\documentclass[twocolumn,showpacs,showkeys,preprintnumbers,amsmath,amssymb,pre]{revtex4}
\usepackage{dcolumn}
\usepackage{bm}
\usepackage{latexsym,epsfig,amssymb}
\usepackage{color}
\usepackage{scalerel}

\DeclareMathOperator{\erfc}{erfc}

\begin{document}

\preprint{\today}

\title{Contact statistics in populations of noninteracting random walkers in two dimensions}

\author{Mark Peter Rast}
\affiliation{Department of Astrophysical and Planetary Sciences, Laboratory for Atmospheric and Space Physics, University of Colorado, Boulder, CO 80309,USA}

\begin{abstract}
The interaction between individuals in biological populations, 
dilute components of chemical systems, or particles transported by turbulent flows depends critically on their contact statistics.  
This work clarifies those statistics under the simplifying assumptions that the underlying motions approximate a Brownian random walk and that the particles can be treated as noninteracting.  
We measure the contact-interval (also called the waiting-time or inter-arrival-time), contact-count, and contact-duration distributions in populations of individuals undergoing noninteracting continuous-space-time random walks on a periodic two-dimensional plane 
(a torus),
as functions of the population number density, walker radius, and random-walk step size.
The contact-interval is exponentially distributed for times longer than the mean-free-collision time but not for times shorter than that, and the contact duration distribution is strongly peak at the ballistic-crossing time for head-on collisions, when the ballistic-crossing time is short compared to the mean step duration.
While successive contacts between individuals are independent, the probability of repeat contact decreases with time after a previous contact.
This leads to a negative duration dependence of the waiting-time interval and over dispersion of the contact-count probability density function for all time intervals.   
The paper demonstrates that for populations of small particles (walker radius small compared to the mean-separation or random-walk step size) the mean-free-collision interval, the ballistic-crossing time, and the random-walk-step duration can be used to construct temporal scalings which allow for common waiting-time, contact-count, and contact-duration distributions across different populations.
Semi-analytic approximations for both the waiting-time and contact-duration distributions are also presented.
   
\end{abstract}

\maketitle

\section{\label{sec1}Introduction}

This paper examines contacts between individual particles each of which is undergoing an independent continuous-space-time random walk on a periodic two-dimensional plane 
(a square with periodic boundary conditions, topologically a torus).   
Contact between particles is defined by particle proximity.  The particles are noninteracting, with their trajectories unperturbed by contact.  We study the contact-interval (also called the waiting-time or inter-arrival-time), contact-count, and contact-duration distributions for any individual in the group, and the dependence of these distributions on particle number density, random-walk step-size, and particle radius (one-half the contact distance).  While a closed-form analytic solution for the distribution of the first contact time between two Brownian particles as a function of their separation is known in one and three dimensions~\citep[e.g.,][]{ 2007gfpp.book.....R, 2010JChPh.133c4105H}, an explicit solution in two-dimensions is not known.  This makes direct calculation of both the minimum first-passage contact time (equivalent to the waiting-time for a given realization of particle separations in a population) and the contact-duration distributions difficult.  We have succeeded in deriving approximate forms for these distributions (Section~\ref{sec4.2} and \ref{sec5}), but we rely largely on numerical simulations of many Brownian particles over many times steps to demonstrate scalings and behaviors as a function of population properties.  
 
The study of random walks, Brownian motion, and first passage processes has a long and rich history~\citep[e.g.,][]{1943RvMP...15....1C, nelson1967dynamical, weiss1994aspects, 2007gfpp.book.....R, peres2010brownian, metzler2014first, annurev-conmatphys2019, Grebenkov_2020}.   Many studies have focused on first-passage probabilities for single particle intersections with targets over a wide range of spatially complex and time varying configurations~\citep[e.g.,][and references therein]{2007gfpp.book.....R, metzler2014first}.  Mean first-passage times for individuals and mean encounter time between walkers on networks is also well studied~\citep[e.g.,][]{2009PhRvE..80c6119S, 2011EPJB...84..691Z, 2021PhRvE.103d2312R}.  Closer to the work presented here are studies addressing the probability of encounter between two walkers in confined domains, analytically in one-dimension~\citep{2011JPhA...44M5005T, 2020PhRvE.102c2118L} and numerically in more than one~\citep{2005Lavine}.  In general, closed-form solutions for encounter statistics between individual random walkers in groups of more than two are difficult to achieve, particularly in two-dimensions~\citep[][]{10.2307/3213665}.  Recent work has made headway in analytically determining the fastest first-passage time of a large number of Brownian particles to a target~\citep{2019JNS....29..461B, 2019JNS....29.2955L, 2020PhRvE.101a2413L} and in numerically assessing the long-time exponential behavior of the distribution of the minimum-time for a group of diffusive walkers to encounter an individual diffusing target as a function of system size in one-, two-, and three-dimensional confined domains~\citep{2020PhRvE.102f2109N}.  Here we focus on contact statistics between individuals in an unconstrained population over all time scales.
      
We find that the ratio of the mean-free-collision time (the mean time between collisions if the particle were to move ballistically (without change in direction) at the random walk step velocity) to the random-walk-step duration is a critical parameter.  The waiting time is exponentially distributed for times longer than the mean-free-collision time but not for shorter times (Section~\ref{sec4.1}).  Contact count distributions are consequently nonPoisson over all time intervals, even in populations with large number densities (Section~\ref{sec4.3}).  
When the mean-free-collision time is used to scale the waiting time, its distribution collapses to a common form for all populations sharing the same particle size (Section~\ref{sec4.1}).  
Waiting-time sensitivity to particle size is captured by the ratio of the ballistic-crossing time (the time it takes for two particles to cross on a straight-line and head-on trajectory) to the random-walk-step duration.  For small particles the contact duration distribution is strongly peaked at this ballistic-crossing time, and the ballistic-crossing time can be used to scale the waiting time (Sections~\ref{sec4.4}) and contact-duration time (Sections~\ref{sec5}) between populations differing in particle size. 

Several length scales (or equivalently time scales as described above) determine contact between individuals in populations:  the step-length taken by the walkers (the correlation length or Lagrangian integral scale for more complex motions), the mean separation or number density of the particles, and the contact distance (particle radius).  It is the relative magnitudes of these that determine the contact statistics observed.  
In the random-walk population models presented here, these quantities are prescribed parameters.  In more complex systems they are constrained by the underlying dynamics of the flow, which determine the particle motions, and the population properties, including the nature of the particle interactions.

Because the particles we consider are noninteracting, our results are most relevant to systems in which 
particle contact yields no change, or a low probability of change, in the particles' motions, systems for which proximity rather than direct contact is critical, or systems in which multiple encounters are required before interaction.  
Some examples include contagion in biological systems~\citep[e.g.][]{Changruenngam2020, norambuena2020, 2020PhRvR...2c3239V},      
chemical systems with low reactivity~\citep[e.g.][]{2004PhBio...1..137A, grebenkov2019, 2020PhRvL.125g8102G}, and aggregation under conditions of uncertain coalescence~\citep[e.g.][]{1999Icar..141..388K, 2016A&A...589A.129H}.
More broadly, this work serves as a simplified baseline for understanding contact statistics in systems with more complex interactive dynamics, such as turbulent flows or crowded populations~\citep[e.g.,][]{2004NJPh....6..116M, 2009PhRvE..79d6314R, polanowski2019, zhao2021}. 
 
\section{\label{sec2}Random Walk Model}

We simulate the motion of a collection of particles undergoing independent continuous-space-time random walks on a periodic two-dimensional plane of unit length in width and height 
(a square with periodic boundary conditions).  
Each successive step of each random walk is taken in a random direction (uniformly distributed in angle between zero and $2\pi$) and has a random length (uniformly distributed between zero and a specified maximum).  
The waiting-time and contact-count results presented here are very likely independent of the step-length distribution employed, so long as that distribution has a finite mean and variance to ensure a Brownian diffusive limit~\citep{Donsker1951}.  
We have checked this empirically for populations of individuals sharing a common and constant step-length.  
One might anticipate that the contact-duration distribution, on the other hand, has greater sensitivity to the step-length distribution 
because contact durations between small particles (smaller than the step size) do not typically extend over diffusive time scales.  
Though we have not yet fully investigated this aspect, we show in Section~\ref{sec5}
that contact durations are dominated by ballistic-crossings, and so expect that they too are 
insensitive to the step-length, so long as the mean step size is greater than the particle size.  

In our random-walk model, the position of each walker on the plane is resolved by numerically advancing their position in small sub-steps, with a small random correction made to the size of the final sub-step to avoid multiple random walk trajectories making directional changes simultaneously, which would otherwise occur due to the discrete nature of the numerical steps 
and is not present in real flows.  
In test cases, the correction has no influence on the results presented in this paper, but for consistency the solutions presented here were all computed employing it.
In short, each random walker takes small equal-length numerical sub-steps (with the exception of the small correction to the last sub-step) in the same direction for a specified total number that is uniformly distributed between zero and the maximum step-length.
Since the walkers effectively move at constant speed, the number of numerical sub-steps taken sets the temporal resolution of steps and thus that of the contact-interval and contact-duration measurements.
Aside from studies focused on temporal resolution checks, $10^4$ numerical sub-steps were taken per mean random-walk step in all simulations. 

In addition to checking that the conclusions drawn in this paper do not depend on the numerical details of how the random walks were constructed (as outlined above), we have also checked that the results depend only very weakly, and then only for very low values of particle number, on the imposed domain periodicity.  Instances of solution sensitivity to periodicity are discussed along with the simulation results in Section~\ref{sec4} below, and to mitigate those, solutions for the lowest number density case were computed not only for the unit square but also for a domain three times as long in each direction.  The results presented in this paper are effectively those for populations of a given number density in an infinite two-dimensional  domain.

The fundamental parameters of the multiple-walker random-walk simulations are the walker radius $a$ and the mean step length $\delta r$, which are taken common to all the walkers in any given simulation, and the number density of walkers in the domain $n$, or equivalently the mean nearest-neighbor separation in the population $0.5/\sqrt{n}$~\citep[][]{2009PhRvE..79d6314R}.  Contact is defined as a separation of $2a$ or less between walkers, with the motions of the individuals unchanged by contact.  No interaction between individual walkers is modeled.  
The particles overlap during times of contact and move along trajectories independent of the contact between them.
This, along with the periodic boundary conditions imposed, ensures that the particle positions, which are initially uniformly distributed on the plane ($x$ and $y$ positions independently and uniformly distributed between zero and one) remain uniformly distributed as the positions evolve. 

In units of the domain width and height, the range of parameter values employed for this paper are:  walker radius $a$ ranging from 0.00005 to 0.035, mean step size $\delta r$ equal to either 0.01255 or 0.0251, and number density $n$ ranging from 4 to 1600 per unit area.
Importantly it is the relative magnitudes of these quantities, not their individual values, that govern the solution statistics.       
      
\section{\label{sec4}Contact interval}

\subsection{\label{sec4.1}Simulation results}

\begin{figure}[t!]
\centerline{\includegraphics[width=6cm,trim=2cm 0cm 2.0cm 0.0cm]{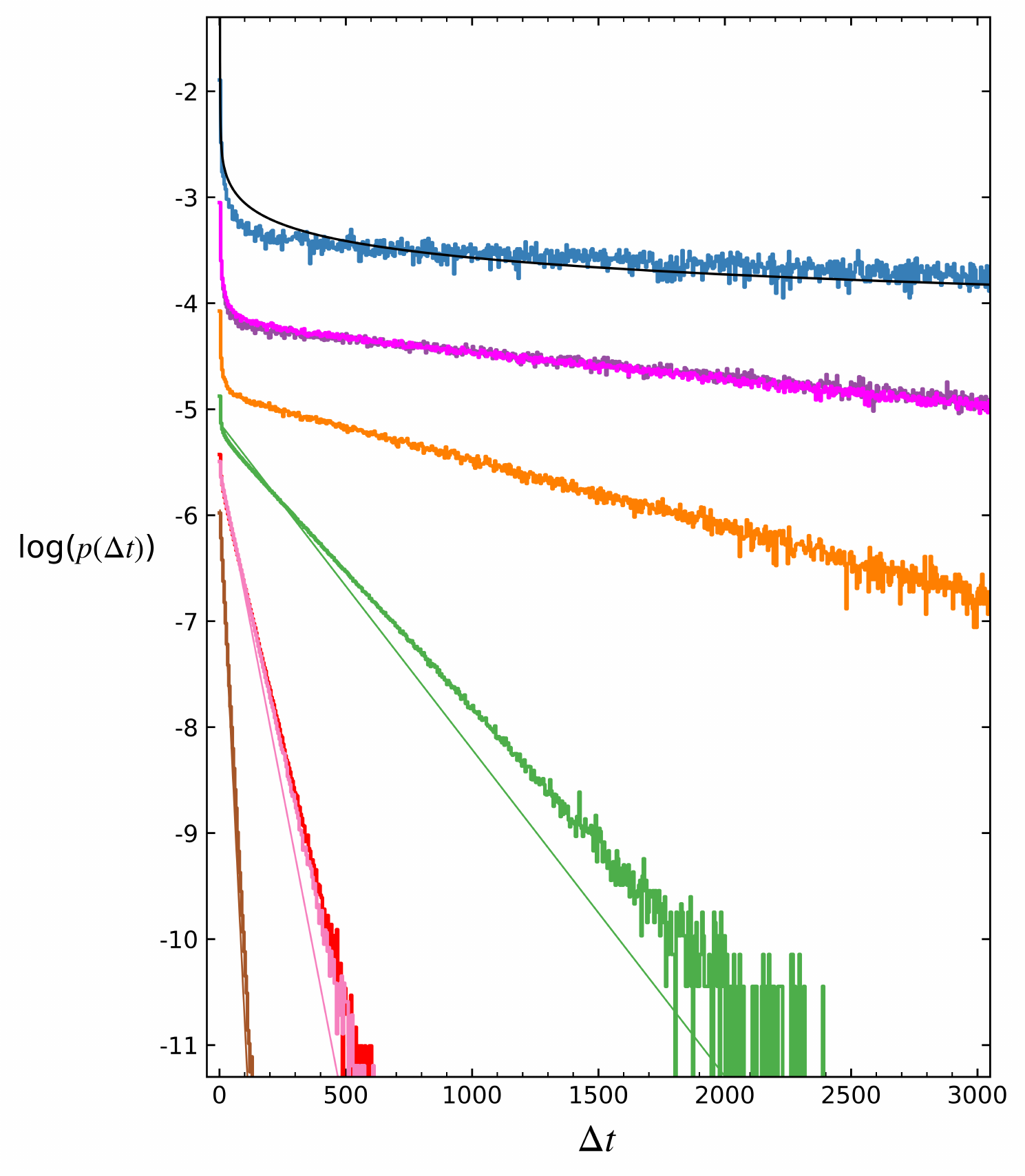}}
\caption{Normalized probability densities $p(\Delta t)$ of the time interval $\Delta t$ between two successive contacts for any individual random walker in a population of walkers.  Time is measured in units of mean step duration.  Distributions below the upper-most one are offset vertically by factors of one-tenth for clarity.  Cases differ only in walker number density ($n=[4, 10, 25, 100, 400, 1600]$ {\it top} to {\it bottom}, {\it blue} to {\it brown}).  They share the same mean step-size ($\delta r=0.0251$) and walker radii ($a=0.0005$).
The nearly indistinguishable overlapping distributions
second from {\it top} are from simulations with 
identical $n$ conducted separately in $1\times1$ and $3\times3$ domains.  These were undertaken to test the waiting-time distribution sensitivity to domain periodicity. Overlapping distributions fifth from the {\it top} result from populations having the same product of $n\,\delta r$ but differing in $n$ and $\delta r$ individually. 
The exponential mean-free-collision interval distribution is over-plotted with {\it solid} lines for the three cases with highest walker number densities (nearly indistiguishable from the actual distribution in the highest number density case (shown {\it brown})).  An analytic approximation (see Section~\ref{sec4.2}) to the low number density waiting time distribution ({\it upper most blue} plot) is indicated with a {\it solid black} curve.}
\label{fig3}
\end{figure}

The interval between contacts $\Delta t$ for each walker in the simulations was computed as the time elapsed between the end of a previous contact (the end of a time period during which the walker was within a distance $2a$ of another walker) to the beginning of the next.  
In all but those cases with the largest walker radii (see Section~\ref{sec5}), the occurrence of simultaneous contact between more than two-individuals is vanishingly rare and so this interval represents the time between binary contacts.  
The normalized waiting-time probability densities $p(\Delta t)$ that result for walkers in a series of simulations differing only in number density are shown in Figure~\ref{fig3}.   The populations simulated for these plots have different walker number densities $n$, but share the same mean step-size $\delta r$ and contact distance $2a$.  
The distributions in the figure are are shifted downward vertically, each by a factor of ten, for clarity, with the uppermost curve ({\it blue}) plotting the unshifted distribution obtained from the simulation with the lowest walker number density ($n=4$) and the lowermost plot ({\it brown}) showing the distribution obtained from the simulation with the highest walker number density ($n=1600$).
Over the remainder of this section we will show that the differences between the waiting-time distributions apparent in Figure~\ref{fig3} are due to the differences in the ratio of a population's mean-free-path length
(the mean distance a walker would travel between collisions if it were to move ballistically at the random-walk-step velocity)
to the actual mean-step length taken by the individual walkers.

For high number densities (short mean-free-path compared to the mean-step length), the walkers take few steps between collisions.  
In this collisional limit, many walkers move on nearly ballistic trajectories between successive contacts and
the interval  between contacts is approximately distributed as the mean-free-collision interval, 
$p(\Delta t)=1/{\bar\tau}\,\exp(-\Delta t/{\bar\tau})$, 
where the mean-free time ${\bar\tau}= 1/(4\,n\,a\,{\bar v})$ (e.g.,~\cite{chapman1990}).  With ${\bar v}$ taken to be the average relative ballistic velocity~\citep[e.g.,][]{2014AmJPh..82..602P}, ${\bar v}=\sqrt{2}\,{v}$, and with time is measured in units of the mean step duration, 
${{\delta t}}=1$, so that $v=\delta r/{{\delta t}} = \delta r$, the mean-free time becomes ${\bar\tau}= 1/\left(4\sqrt{2}\,n\,a\,\delta r\right)$.  
The  limiting exponential collision probability density 
is over-plotted with {\it solid} lines in Figure~\ref{fig3} for the three cases with highest walker number densities.  
It is nearly indistinguishable from the actual distribution for the $n=1600$ case (shown {\it brown}).
As $\bar\tau \rightarrow 0$ (because $n\,\delta r \rightarrow\infty$ for a given finite particle radius), the range of waiting times over which distribution is non-exponential goes to zero and the distribution becomes strictly exponential for all waiting times~\citep{10.2307/3213665}.  In that limit, all walker contacts occur on strictly ballistic trajectories.
Note that for $\bar\tau<1$ the timescale over which the collision-interval distribution becomes non-exponential becomes less than one-mean step time and the step-duration (or equivalently the step-length) distribution itself then becomes important to the solution. 

With time measured in units of the mean step duration, the ratio of the mean-free-collision time to mean step duration is given by $R=1/\left(4\sqrt{2}\,n\,a\,\delta r\right)$, with $R\approx[3520, 1410, 563, 141, 35.2, 8.80]$ for the cases shown {\it top} to {\it bottom} in Figure~\ref{fig3}.
While the populations we simulate fall short of the strictly ballistic limit,
the waiting-time distributions approach the collisional exponential over all but the very shortest time intervals in the highest number density (lowest $R$) cases.
  
In the opposite large $R$ limit, low number densities or small step lengths (long mean-free-path compared to the mean step-length), the walkers undertake many random walk steps between contacts.  In this Brownian limit, the waiting time for any individual in the population is the minimum statistic of the first-passage time to a separation of $2a$ with any other individual.  An  approximate semi-analytic solution for its form is developed in the next section (\ref{sec4.2}) and is over-plotted with a thick black curve for the lowest number density case in Figure~\ref{fig3}.  It does a reasonable job of capturing the distribution over these time scales.

\begin{figure}[t!]
\centerline{\includegraphics[width=6.25cm,trim=2cm 0cm 2cm 0cm]{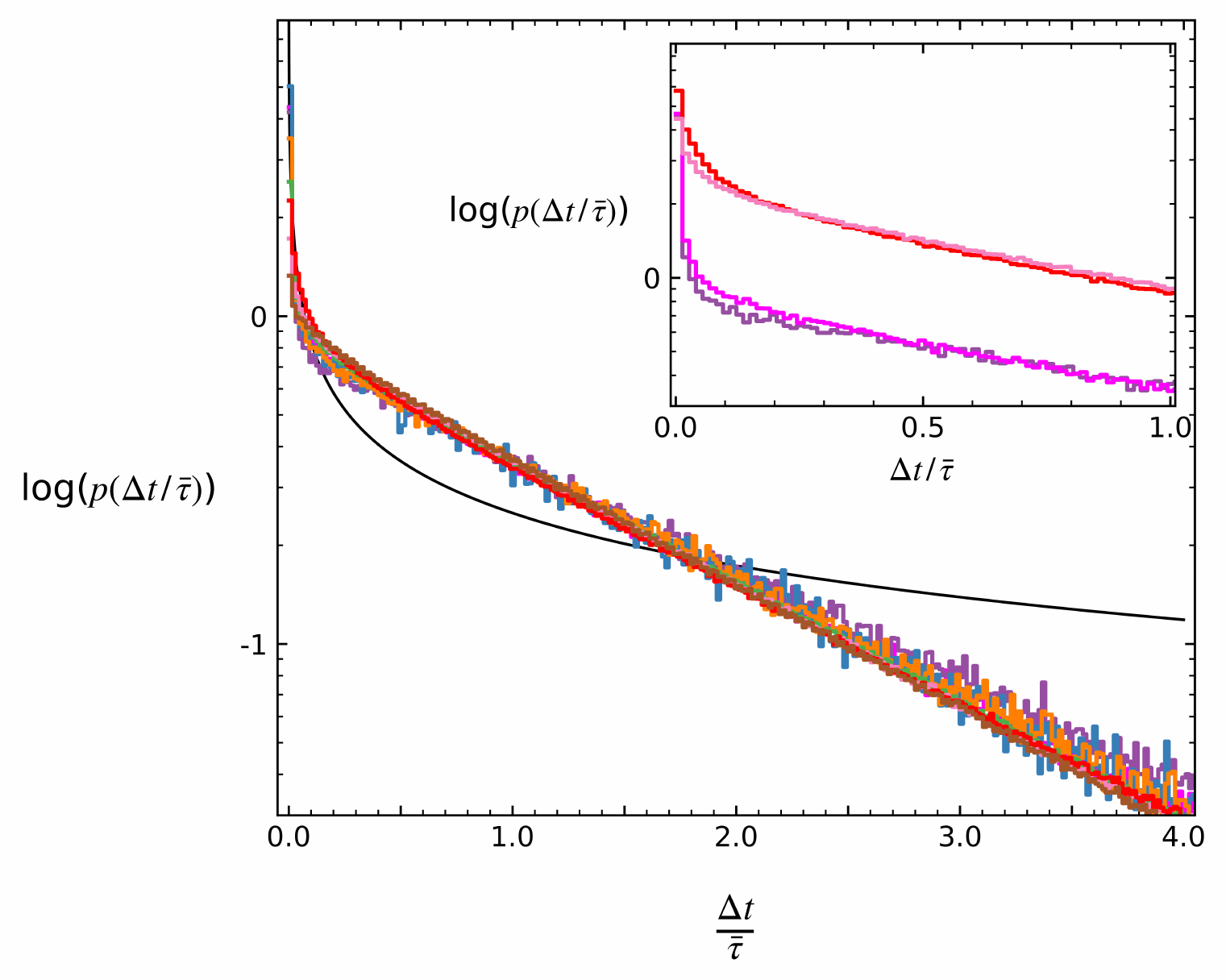}}
\caption{Probability density $p(\Delta t /{\bar\tau})$ of the waiting time $\Delta t$ (with time is measured in units of mean step duration) scaled by the mean-free-collision interval ${\bar\tau}= 1/(4\,n\,a\,{\bar v})$, with relative velocity ${\bar v}=\sqrt{2}\,\delta r/{\delta t}$.
Cases differ in walker number density $n=[4, 10, 25, 100, 400, 1600]$, sharing the same mean step-size $\delta r=0.0251$ and walker radii $a=0.0005$.  Color scheme and underlying simulations match those of Figure~\ref{fig3}, but here are difficult to distinguish after the $\Delta t$ scaling.  An analytic approximation (Section~\ref{sec4.2}) to the waiting time distribution is indicated with an underlying {\it solid black} curve.
Note that while plotted over the full abscissal range this solution is only valid for times very short compared to $\bar\tau$.
The inset replicates the plots for two pairs of random walk simulations.  The lower pair illustrates the dependence on domain periodicity at small number densitites, and the upper pair (offset vertically) displays the small mismatch between simulations with the same value of $n\,\delta r$ but with differing number density $n$ and mean step length $\delta r$ individually.}
\label{fig5}
\end{figure}

All the simulations illustrated by Figure~\ref{fig3} share the same random-walker radius $a$, thus each distribution shown approximates a family of distributions sharing the same value of $n\,\delta r$.  The overlapping distributions plotted fifth from the top in Figure~\ref{fig3} illustrate this with simulations whose individual $n$ and $\delta r$ values differ by  factors of 2 and 1/2.  While distributions are nearly identical, small differences are apparent 
(barely distinguishable here, but see Figure~\ref{fig5} inset).  
The differences reflect the unaccounted for changes in the relative amplitudes of the mean nearest-neighbor particle separation and the mean step size to the particle radii as $n$ and $\delta r$ are changed.  The simulations are not strictly similar when the particle size is held constant.  There are weak sensitives to the change in the relationships between the contact distance and the step-size and between the contact distant and the mean spacing of the walkers in the domain, because at close range and over short waiting times the frequency of walker contacts is sensitive to the walker size (see Section~\ref{sec4.4}).

Separately, sensitivity to domain periodicity was investigated.  The pair of waiting time distributions plotted second from the top in Figure~\ref{fig3} result from two random walk simulations conducted in different size domains ($3\times3$ shown {\it magenta} and $1\times 1$, shown {\it purple}) but otherwise sharing identical parameter values (identical values of $n$, $a$, and $\delta r$).  
While nearly indistinguishable in this plot, there are small differences in shape of the two distributions that are more apparent in the inset of Figure~\ref{fig5}.
This very weak sensitivity to domain periodicity decreases even further with increasing domain size and/or particle number density because contacts that result from periodic edge crossings become less numerous relative to those occurring in the bulk of the domain as the total number of particles simulated increases.
Thus, the boundary periodicity we impose plays no significant role in the results we present.

The importance of the mean-free-collision time as a scaling parameter is more clearly illustrated by Figure~\ref{fig5} which re-plots Figure~\ref{fig3} after scaling the waiting time by $\bar\tau$ (with no vertical offset of the distributions).  Application of this scaling means that the abscissal extent in Figure~\ref{fig5} is about 4.7 times that shown in Figure~\ref{fig3} for the $n=4$ distribution and about one-percent of that shown in Figure~\ref{fig3} for the $n=1600$ distribution.  When scaled by the mean-free-collision time, the waiting-time probability densities collapse to a nearly common distribution.  The small differences remaining are due to the walker-size sensitivities and domain periodicity effects discussed above, and illustrated by the figure inset.  

\subsection{\label{sec4.2}Approximate solution in the Brownian limit}

In a population characterized by large $R$ (long mean free-path compared to the mean step-length), particles undertake many random walk steps between contacts, either because the step length is very short or the number density is very low.  
In this Brownian limit, the waiting time for any individual is the minimum order statistic of the first-passage time for any other member of the population to a circle of radius of $2a$ surrounding the target (accounting for relative velocities in the diffusion coefficient), and the waiting-time distribution is determined by evaluating the minimum statistic given the pair-separation distribution on the plane  (see e.g.,~\cite{gumbel1962, davidandnagaraja2003} for general expositions of order statistics, 
and~\cite{lawley2020, 2020NJPh...22j3004G} for recent application to the fastest first-passage time to a target). 
Unfortunately, the distribution of the first-passage time to a circular object is not explicitly known in closed form in two dimensions, 
and so the minimum statistic is not easily evaluated.

Equivalently, in terms of pair separation, the waiting time for any individual is the minimum statistic of the 
first-passage time to a pair separation of $2a$ with any other individual in the population. 
Pair-separation studies have a rich history, particularly in turbulent transport~\citep[e.g.,][and references therein]{2001AnRFM..33..289S, 2006Sci...311..835B, 2009AnRFM..41..405S, 2011PhRvL.107u4501R, 2015JFM...772..678B, 2016PhRvE..93d3120R}, 
where the focus is often on mixing and thus dispersion from small to large separations.  We 
outline in Appendix~\ref{appendix0} some important characteristics of the time-reverse problem,  
from large separations to small, that underlie the contact considerations here.
Importantly, while the distribution of the distance between individual pairs as a function of time is well known in the Brownian limit (Equation~\ref{eqn2} below), even in this limit the distance between any two walkers only approximates a one-dimensional diffusive process at short and long times (Appendix~\ref{appendix0}).  Over intermediate times the pair-separation variance is a nonlinear function of time, making an exact closed-form expression for the first-passage time over all time scales difficult.  We develop here an approximate solution for short first contact times in large $R$ populations (Figure~\ref{fig3})
or equivalently for all populations over sufficiently short times compared to the mean-free-collision interval (Figure~\ref{fig5}).

In the Brownian limit, the distance $r_s$ between two unbiased random walkers, as a function of their initial separation $r_0$ at $t=t_0$, is Rice distributed~\citep[e.g.,][]{1945BSTJ...24...46R,cole2010heat,2018ApJ...854..118A}
\begin{equation}
p(r_s\vert r_0)  = {{r_s}\over{\sigma^2}}\exp\left(-{{r_s^2+r_0^2}\over{2\sigma^2}}\right)I_0\left({{r_0r_s}\over{\sigma^2}}\right)\ ,
\label{eqn2}
\end{equation}
where $I_0$ denotes the lowest order modified Bessel function of the first kind~\cite{1972NBSAbramowitzandStegun}, $r_0$ is the initial separation at $t=t_0$, and $\sigma^2= 4\,D\,(t-t_0)$ with $D=\delta r^2 /4 \delta t$.  
Unfortunately this distribution does not readily yield a closed form expression for the first-passage-time probability density, and thus, can not be conveniently used to determine the waiting time distribution, the minimum order statistic of the first-passage time to a given separation.
However, Brownian pair separation (Equation~\ref{eqn2}) has two limiting forms (Appendix~\ref{appendix0}).  
For $\delta t\lesssim t-t_0\leq r_0\,\delta t/(2\delta r)$, where $r_0\,\delta t/(2\delta r)$ is the earliest possible time that the distance between two walkers can equal zero, the pair-separation distribution evolves as a truncated Gaussian.
While the Gaussian distribution is convenient for the first-passage time calculation, it is not relevant.  The first-passage time distribution for any individual walker can not be determined using the pair-separation distribution valid for times less than or equal to the minimum time for contact.

In the opposite limit, 
$t-t_0\gtrsim4 r_0^2\delta t/\delta r^2$, many steps have been taken and the pair-separation distribution is approximately Rayleigh (Appendix~\ref{appendix0}), 
\begin{equation}
\label{eqn6}
p(r_s,t\vert r_0,t_0)\approx{{r_s}\over{\sigma^2}}\ \exp\left(-{{(r_s-r_0)^2}\over{2\sigma^2}}\right)\ .
\end{equation}
Using this pair-separation distribution to evaluate the probability density of first-passage to separation of $2a$ yields its Laplace transform~\citep[e.g.,][]{2007gfpp.book.....R}
\begin{equation}
{\tilde f}(s\vert r_0)={{K_0\left(\!{\sqrt{{(r_s-r_0)^2\,s}\over{2D}}}\right)}\over{K_0\left(\!{\sqrt{{(r_s-2a)^2\,s}\over{2D}}}\right)}}\ ,
\label{eqn7}
\end{equation} 
with the Bromwich integral solution
\citep[see e.g.,][]{1959chs..book.....C},
\begin{equation}
\begin{split}
f(t\vert r_0,t_0)={2\over{\pi}}\int_0^\infty \!\!&u\,e^{-u^2(t-t_0)}\ \\
&{{J_0(Bu)Y_0(Au)-J_0(Au)Y_0(Bu)}\over{\left[J_0(Bu)\right]^2+\left[Y_0(Bu)\right]^2}}\ du\ ,
\end{split}
\label{eqn8}
\end{equation}
where $A=\vert r_s-r_0\vert/{\scriptstyle \sqrt{2D}}$, $B=\vert r_s-2a\vert/{\scriptstyle \sqrt{2D}}$, and 
$K_0$, $J_0$ and $Y_0$ denote the lowest order modified Bessel function of the second kind and the lowest order Bessel functions of
the first and second kind respectively~\cite{1972NBSAbramowitzandStegun}.
The factor $2D$ in $A$ and $B$ results because both particles are moving with the same Brownian properties and thus the effective diffusion coefficient is doubled.

Beyond this formal solution, the inverse Laplace transform of Equation~\ref{eqn7} can be determined analytically only in the large $s$ limit. This corresponds to short first contact times (small $t-t_0$).  In this limit and to lowest order, $K_0(x)\approx{\scriptstyle{\sqrt{\pi/2}}}\ \ x^{\scriptscriptstyle{-{1/2}}}\ e^{-x}$~\cite{1972NBSAbramowitzandStegun},
and the inverse transform of Equation~\ref{eqn7} is
\begin{equation}
f(t\vert r_0,t_0)\approx{{2a}\over{{r_0}}}{{(r_0-2a)}\over{\sqrt{8\pi D\,(t-t_0)^3}}}\exp\left(-{{(r_0-2a)^2}\over{8\,D\,(t-t_0)}}\right)\ ,
\label{eqn9}
\end{equation}
when taken to be independent of $r_s$, as appropriate.  The first contact time $t-t_0$ between any two walkers is L\'{e}vy distributed, and depends on their initial separation $r_0$ and radii $a$.  

The approximations made in the development of this first-passage-time distribution imply that it is valid only for short first contact times between walkers that have taken many steps before contact.  
Within our context, this is the short-time limit of populations characterized by large $R$ (long mean-free time to mean step duration).  
It is applicable as well to very short waiting times in any population (Figure~\ref{fig5}), 
because, for times very short compared to the mean-free-collision interval, the waiting time is dominated by particles who have made directional changes while still in close proximity following a previous contact (Sections~\ref{sec4.3} and~\ref{sec4.4} below).    
We note the curious fact that, with these approximations, the first-crossing time distribution for a given $r_0$ in two dimensions is equivalent to that without approximation in the three-dimensions 
\citep[e.g.,][]{2007gfpp.book.....R}.  
This does not mean that the waiting time distributions derived using it will be then same as that in three dimensions because the distributions of particle separations sampled 
by the populations in two and three dimensions are different.   

Given the first-passage-time probability density (Equation~\ref{eqn9}), one can calculate the waiting time for a given distribution of particle separations as the minimum order statistic.  
The waiting time for any individual in a large collection of Brownian particles is the minimum first-passage time to a separation of $2a$ from a value larger than this.  It is the shortest time for the separation between a particle and any other member of the population to reach a value of  $2a$.  
For independent but non-identically distributed variates (as is the case for first-passage-time distributions for each member of the population since each depends on a different separation distance from the target individual), 
the minimum oder-statistic is given by $F_{(1)}(x)=1-\prod_i^M \left(1-F_i(x)\right)$,
where $\prod$ indicates the product and $F_i(x)$ are the cumulative distributions from which $M$ samples are drawn (e.g.,~\cite{caoandwest1997,davidandnagaraja2003}).  For the approximate
L\'{e}vy distributed first-crossing time (Equation~\ref{eqn9}), the cumulative distributions are complementary error functions~\cite{1972NBSAbramowitzandStegun}, so
\begin{equation}
F_{(1)}(\Delta t)\approx1-\prod_i^{N-1}\left[1-{{2a}\over{{r_0}_i}}\erfc\left(\sqrt{{{({r_0}_i-2a)^2}\over{8D\Delta t}}}\right)\right]   \ ,
\label{eqn10}
\end{equation}
where $\Delta t=t-t_0$ is the waiting time interval for an individual walker and ${r_0}_i$ is the separation between it and each of the $N-1$ other walkers on the plane.  

Since each Brownian particle in the population isotropically and randomly samples the square plane
(as ensured by the periodic boundary conditions imposed), the distances between them 
at any instance in time are distributed as the distances between two randomly chosen points on the unit square.
This is given by~\cite[][]{jphilip2007}, 
\begin{equation}
p(r_0) =\begin{cases}
2\pi r_0-8r_0^2+2r_0^3\ , \ \ \ \ \ \ \ \ r_0\le 1 \\
\\
\begin{split}
-(2\pi+&4)r_0+8r_0\sqrt{r_0^2-1}-2r_0^3 \\
&+8r_0\arcsin\left({1\over{r_0}}\right) \ , \ \ \ \ \, 1<r_0\le\sqrt{2} \ .
\end{split}
\end{cases}
\label{eqn11}
\end{equation}
The cumulative waiting-time distribution $F_{(1)}(\Delta t)$ is then determined by sampling
the planar-separation distribution (Equation~\ref{eqn11}) for ${r_0}_i$,
$N-1$ times, and evaluating Equation~\ref{eqn10} with those values.  The average cumulative distribution is achieved by repeating this process many times, and the corresponding average waiting-time probability density is plotted as a {\it black} curve in 
Figures~\ref{fig3} and~\ref{fig5}.  It does a reasonable job of approximating the distributions observed for populations with low number density, large mean-free path and short step sizes, or over time intervals very short compared to the mean-free-collision time in all populations.

\subsection{\label{sec4.3}Contact counts}

\begin{figure}[t!]
\centerline{\includegraphics[width=6.25cm,trim=2cm 0cm 2cm 0cm]{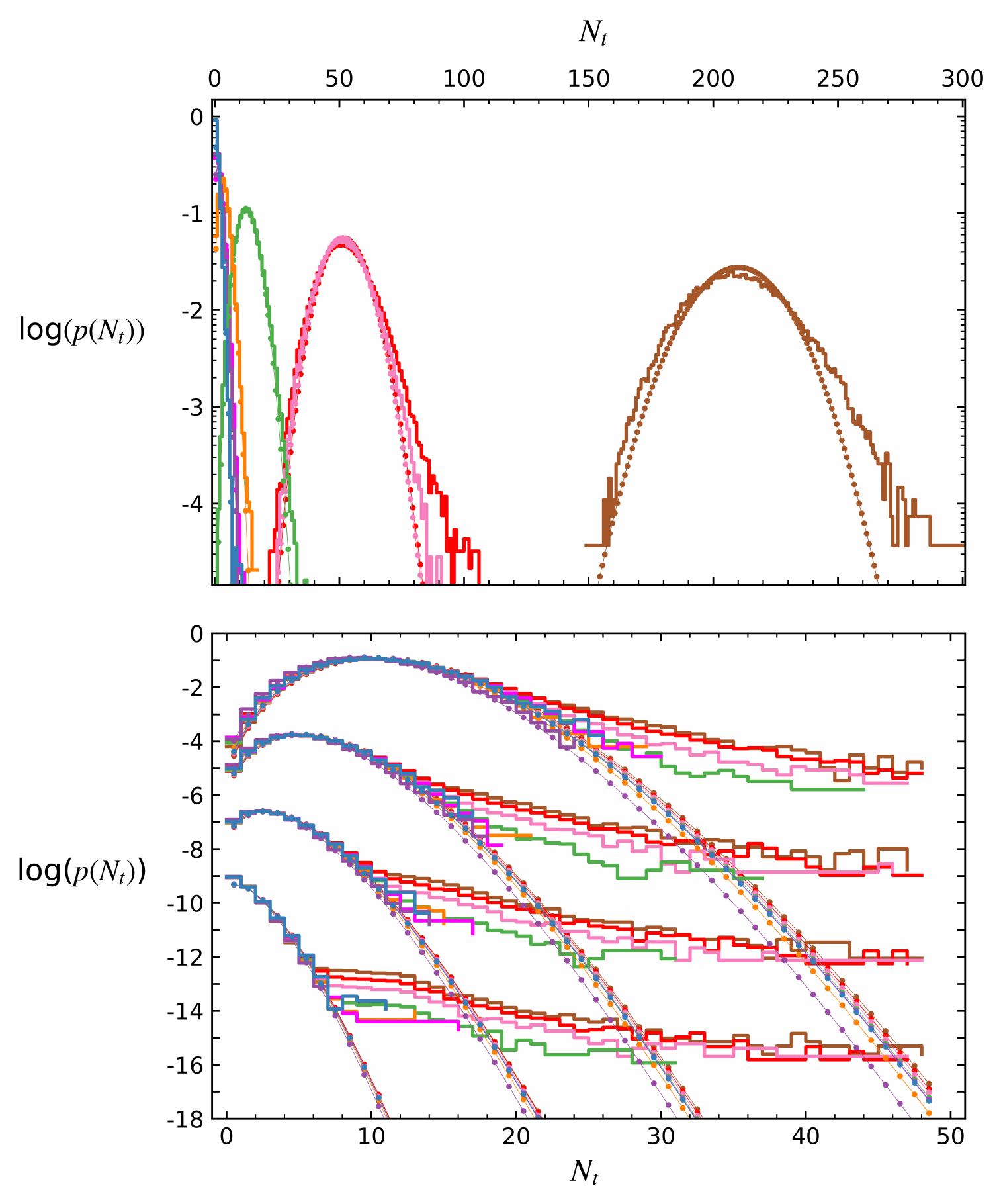}}
\caption{Probability density $p(N_t)$ of the number of contacts $N_t$ experienced by a walker in time interval $t$.  Simulation runs and color coding are the same as those of Figures~\ref{fig3} and~\ref{fig5}, 
with cases differing in the walker number density $n=[4, 10, 25, 100, 400, 1600]$ (shown {\it blue} and {\it brown}, {\it left} to {\it right} in upper panel) but sharing the same mean step-size $\delta r=0.0251$ and walker radii $a=0.0005$.  
In the upper panel, count distributions are plotted for an interval of 2000 steps.  In the lower panel, count distributions are plotted for  
four different time intervals scaled by the mean-free-collision time:  approximately 11, 5.7, 2.8, and 0.57$\,\bar\tau$ (offset {\it top} to {\it bottom}).  For reference these correspond to 40000, 20000, 10000, and 2000 random-walk steps in the $n=4$ simulation (shown {\it blue} and {\it purple} in $3\times3$ and $1\times1$ domains respectively) and 100, 50, 25, and 5 steps in the $n=1600$ run (shown {\it brown}).  Poison distributions based on the mean $N_t$ values are over plotted with {\it solid-dotted} curves in both panels.}
\label{fig6}
\end{figure}

The non-exponential waiting time distributions observed imply non-Poisson count statistics (e.g.,~\cite{taboga2012lectures}).  The rapid decline in probability density at short waiting times yields a rapid decrease in the hazard function $h(\Delta t)=p(\Delta(t)/\left(1-P(\Delta t)\right)$, where $P(\Delta t)$ is the cumulative waiting time distribution, with increasing waiting time.  This suggests negative duration dependence and over-dispersion of the contact count probability mass function (e.g.,~\cite{winkelmann1995}), $p(N_t)$, where $N_t$ is the number of contacts an individual walker has with any other over a time interval of $t$ steps.  Contacts between random walkers are independent, but the probability of contact between any two individuals depends on the elapsed time since the last occurrence.  It is more likely for two random walkers recently in contact to contact each other again because the mean and variance of their separation increases with time.  In a confined space recontact depends on the domain size~\citep{2005Lavine}.  For a population of individuals on an effectively infinite plane 
(a periodic domain of sufficient size, so that the number of contacts due to periodic edge crossings is negligible compared to those occurring in the bulk of the domain, as discussed in Section~\ref{sec4.1} above), 
the importance of recontact compared to new contact depends on the value of the mean-free collision time.  For times shorter than the mean-free collision time, the waiting time is not 
exponentially distributed because the likelihood of recontact with a previously contacted walker is greater than the likelihood of contact with a new walker.

The upper panel of Figure~\ref{fig6} displays (for each of the simulations whose waiting time distribution is shown in Figures~\ref{fig3} and \ref{fig5}) the number of contacts an individual walker experiences over a 2000 step interval.  Unsurprisingly, 
higher counts occur in simulations with higher walker number densities (smaller $R$).   
Perhaps less expected, is that significantly non-Poisson count statistics, over-dispersion of the contact count probability mass function, are apparent even at high number densities and even in the cases for which the waiting time distribution is exponential 
at all but the very shortest waiting times.  The short waiting time excess results in non-Poison count statistics in all the simulations.  

Scaling the count intervals by the mean-free-collision time collapses the count statistics as 
it did for the waiting time distributions.  
In the lower panel of Figure~\ref{fig6}, the count distributions are plotted in groups for all the simulations using 
four different time intervals each scaled by the mean-free-collision time:  approximately 11$\,\bar\tau$, 5.7$\,\bar\tau$, 2.8$\,\bar\tau$, and 0.57$\,\bar\tau$ (offset {\it top} to {\it bottom}).  
The Poisson distributed cores of the count distributions for these mean-free-collision weighted time intervals overlap, with  
all distributions showing similar non-Poisson count excesses for large count values. 
Somewhat unexpectedly, over dispersion of the distribution appears to be slightly greater 
in small $R$ simulations (high number density) than in large $R$ simulations.  
This seems counter to the expectation that as $\bar\tau \rightarrow 0$ (as $n\rightarrow\infty$) all walker contacts should occur on strictly ballistic trajectories and that therefore the negative duration dependence of the waiting time should vanish because there is no chance of recontacting a previously contacted walker before another.  The unexpected behavior is due to the finite particle size (interaction distance).  
As the particle size (along with step length) are fixed across these simulations, 
the particles are bigger relative to the inter-particle spacing as the number density
increases. This enhances the relative importance of recontact with previously contacted particles compared to new contact even after the count interval has been scaled to account for the differing $n\,\delta r$ values. 

\subsection{\label{sec4.4}Sensitivity to interaction distance}

The contact interval between walkers at close range is sensitive to the interaction distance (or equivalently the walker radii).  As the particle radii increase, toward the step length, recontact becomes more probable because smaller changes in direction are required for recontact.  In the extreme, when the radius approaches the mean nearest-neighbor separation, multi-walker overlap becomes much more likely.
The mean-free-collision time scaling we have employed to this point does not capture particle size effects.

\begin{figure}[t!]
\centerline{\includegraphics[width=6.25cm,trim=2cm 0cm 2cm 0.0cm]{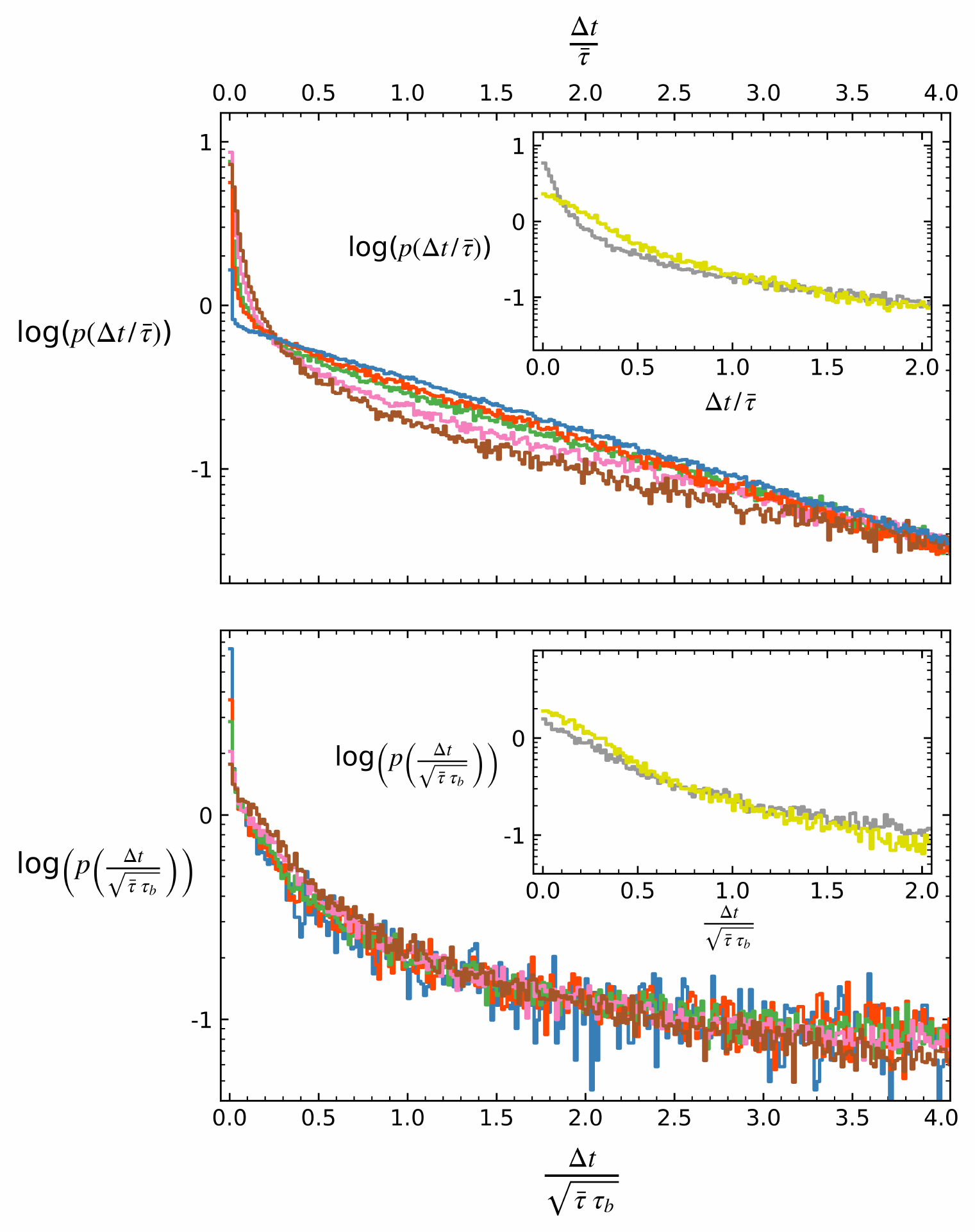}}
\caption{Waiting-time distributions for simulations that share the same step size and number density but differ in particle size (interaction distance). Colors indicate distributions ({\it top} to {\it bottom}, {\it blue} to {\it brown}) obtained from simulations with walker radii $a\approx[0.004, 0.04, 0.08, 0.2, 0.4]\,\delta r$ or equivalently about $[0.00071, 0.0071, 0.014, 0.035, 0.071]$ times the mean nearest-neighbor distance between walkers.  Inset shows waiting-time distributions for simulations with  $a\approx[0.6, 2.8]\,\delta r$ or about $[0.1, 0.5]$ times the mean nearest-neighbor distance, {\it grey} and {\it yellow} respectively.  Waiting time in the upper panel is scaled by the mean-free-collision time $\bar\tau$ and in the lower panel by $\sqrt{\bar\tau\,\tau_b}$, where $\tau_b$ is the ballistic-crossing time for head-on collisions.  The later scaling is independent of $a$, but does not account for multi particle overlap ({\it grey} and {\it yellow} in inset).}
\label{fig7}
\end{figure}

Plotted in the upper panel of Figure~\ref{fig7} are the mean-free scaled waiting time distributions realized in simulations computed with identical walker step size and number densities, but with differing interaction distances.  The {\it orange} curve plots the same waiting-time distribution as that of the same color in Figures~\ref{fig3}~--~\ref{fig6}.  In that simulation the particle radii were equal to 4\% of the mean step size and 0.7\% the mean nearest-neighbor distance.  Walker radii in the remaining simulations illustrated range from one-tenth ({\it blue} curve) to seventy times ({\it yellow} curve in the inset) those values.
As walker radii get larger, the non-exponential behavior of distribution at short waiting times becomes more pronounced even when the waiting time interval is scaled by $\bar\tau$.  
Short waiting times become more probable as the walker radius (interaction distance) approaches the step-size and as it approaches the mean nearest-neighbor separation, multi-walker overlap becomes more likely ({\it grey} and {\it yellow} curves, Figure~\ref{fig7} inset).  
In the opposite limit, very small particle size relegates the non-exponential behavior to very short times.  
As $a \rightarrow 0$ for finite $n\,\delta r$, the time over which the walkers are in close enough proximity for non-ballistic motions to be important for recontact also goes to zero, and as this limit is approached, the waiting time distribution becomes exponential for all but the shortest time ({\it blue} curve, Figure~\ref{fig7}).  

A relevant time scale for contact between walkers is the ballistic-crossing time for head-on collisions, $\tau_b=2a/\delta r$, with time measured in mean step duration so that $\delta t=1$, and a walker-radius independent time scale can be constructed from the product ${\bar\tau\,\tau_b}=1/(2\sqrt{2}\,n\,\delta r^2)$.  Simulations sharing the same value of $n\,\delta r^2$ but with differing waker radii $a$ yield overlapping waiting-time distributions when the waiting time is scaled by  $\sqrt{\bar\tau\,\tau_b}$ (lower panel Figure~\ref{fig7}).  Deviations from this scaled waiting-time distribution occur when $a$ approaches the step size or the mean nearest-neighbor distance ({\it brown, grey, {\rm and} yellow} curves). 

As pointed-out by anonymous referees, a number of subtleties remain.  Apart from repeat encounters between nearby walkers, the non-ballistic nature of the particle trajectories themselves introduces non-Poisson behavior.  Changes in direction of a finite size particle causes overlap of the area explored by that particle before and after the directional change.  This implies non-constant contact rates with other walkers (via a reduction in the new area explored per unit time immediately after each directional change), and thus departures from Poisson distributed contact counts which are larger for larger particles.  This effect may be more apparent in populations of walkers with larger radii and lower number densities than those studied here.  Moreover, because of the importance of recontact in our simulations (the high probability of repeated contact very soon after the first because of the particles' close proximity, Section~\ref{sec4.3}), the waiting time distribution for the first contact and that for subsequent contacts may be considerably different.  Characterization of the $n^{\rm th}$ contact waiting-time distribution and its dependence on population number density, random-walk step length, and particle size is of significant interest, particularly under circumstances of low interaction 
probability or when multiple contact encounters are critical for interaction.

\section{\label{sec5}Contact duration}

\begin{figure}[t!]
\centerline{\includegraphics[width=6.25cm,trim=2cm 0cm 2cm 0.5cm]{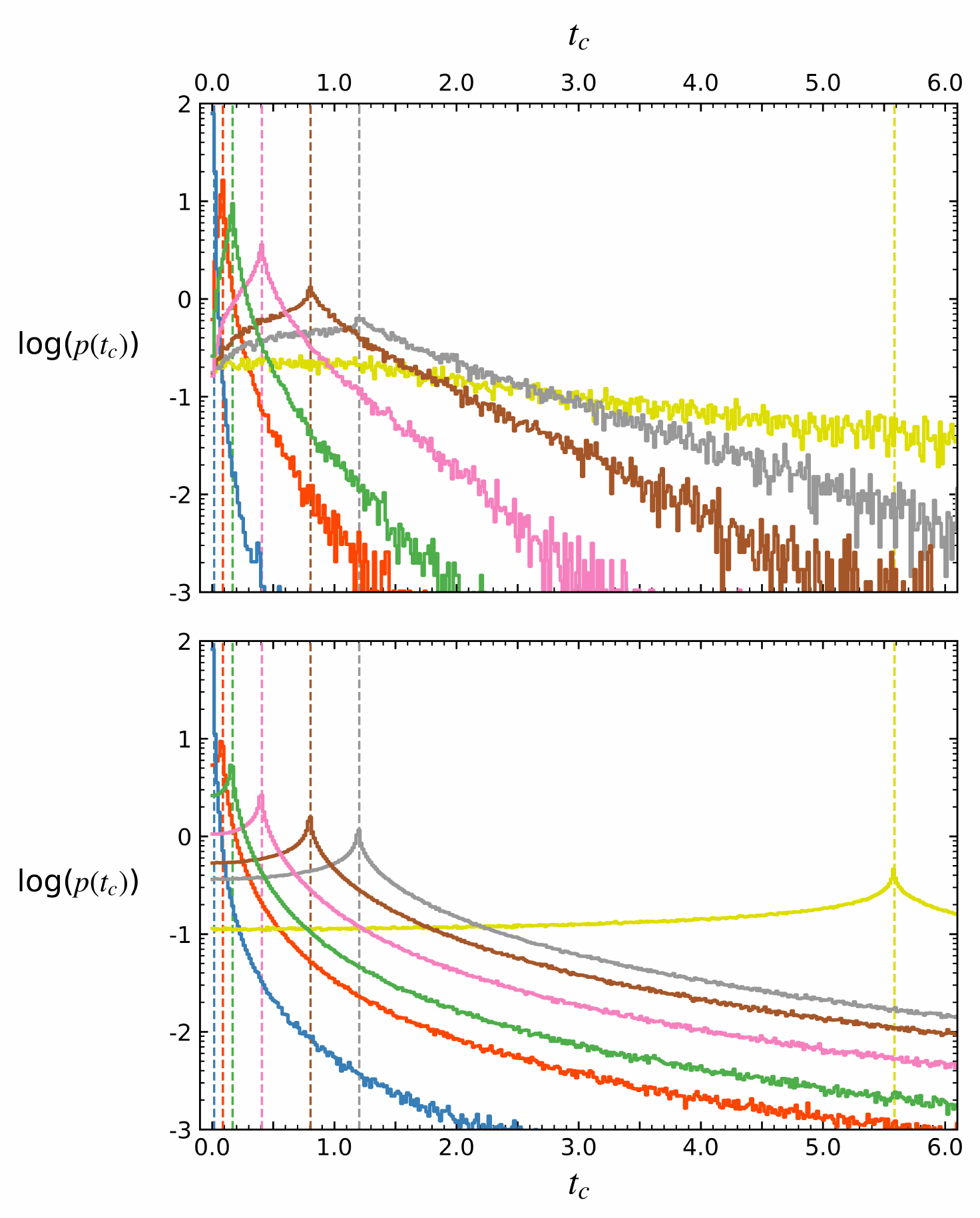}}
\caption{Contact duration $t_c$ (in units of the mean step time) distributions (upper panel) for simulations differing in particle size only. 
Colors indicate distributions ({\it left} to {\it right}, {\it blue} to {\it yellow}) obtained from simulations with walker radii $a\approx[0.004, 0.04, 0.08, 0.2, 0.4, 0.6, 2.8]\,\delta r$ or equivalently about $[0.00071, 0.0071, 0.014, 0.035, 0.071, 0.1, 0.5]$ times the mean nearest-neighbor distance between walkers
(the same simulations and the same color scheme as Figure~\ref{fig7}).  Contact duration distribution for randomly oriented ballistic intersections (lower panel) between objects of the same size as the walkers in the simulations yielding the distributions in the upper panel.  Vertical fiducial lines indicate the ballistic-crossing times for head-on collisions, $\tau_b={{2a}/{\delta r}}$.
Note that, as discussed in the main text, the distributions in the lower panel collapse identically into one when the contact duration is rescaled by $\tau_b$.  Scaled simulation distributions are shown in Figure~\ref{fig9}.}
\label{fig8}
\end{figure}

The contact duration $t_c$ is also sensitive to the walker radius (interaction distance).  The contact-duration probability density is shown in the upper panel of Figure~\ref{fig8} for the same set of simulations as those for which the waiting-time distributions are plotted in Figure~\ref{fig7}.  As expected, the mean contact duration increases with increasing walker size, but additionally the distributions are structured, showing a discrete peak at $\tau_b$. 

In most of the simulations studied, the mean random-walk step size was taken to be much larger than the particle size, which defines the contact distance.  For example, in the simulations in studied in Section~\ref{sec4}, $2a/\delta r\approx0.04$ or $0.08$.   In those underlying the distributions in main body of Figure~\ref{fig7}, $2a/\delta r\approx[0.008, 0.08, 0.16, 0.4, 0.8]$, and in those contributing to the distributions plotted in the inset of Figure~\ref{fig7}, $2a/\delta r\approx[1.2, 5.6]$.  
When the particle radii are smaller than the random-walk step size, most contacts occur over a single step and the contact duration is dominated by the ballistic crossing of the two particles in random directions.  When this is the case, the separation of any two walkers at close range can be approximated by the ballistic equation of motion based on their relative velocity,  
\begin{equation}
\label{eqn4}
\begin{split}
r_s^2=r_0^2+2\,r_0\,\delta r\,t\,(\cos\theta_1&-\cos\theta_2) \\
&-2\,\delta r^2\,t^2\,\cos(\theta_1-\theta_2)\ , 
\end{split}
\end{equation}
where $\theta_1$ and $\theta_2$ are the angles the walker velocity vectors make with the axis between them and time is measured in steps so that their speeds are equal to $\delta r$ ($v=\delta r/\delta t$ with $\delta t =1$).  
The contact duration, in this limit, is the time it takes for the distance between two objects on ballistic trajectories to go from a value of $2a$ (first contact) to less than $2a$ and back (last contact).  It is given by the $t_c > 0$ solutions to
\begin{equation}
t_c={{2a}\over{\delta r}}\,{{\left(\cos\theta_1-\cos\theta_2\right)}\over{1-\cos(\theta_1-\theta_2)}} \ .
\label{eqn12}
\end{equation}

\begin{figure}[t!]
\centerline{\includegraphics[width=6.25cm,trim=2cm 0cm 2cm 0.0cm]{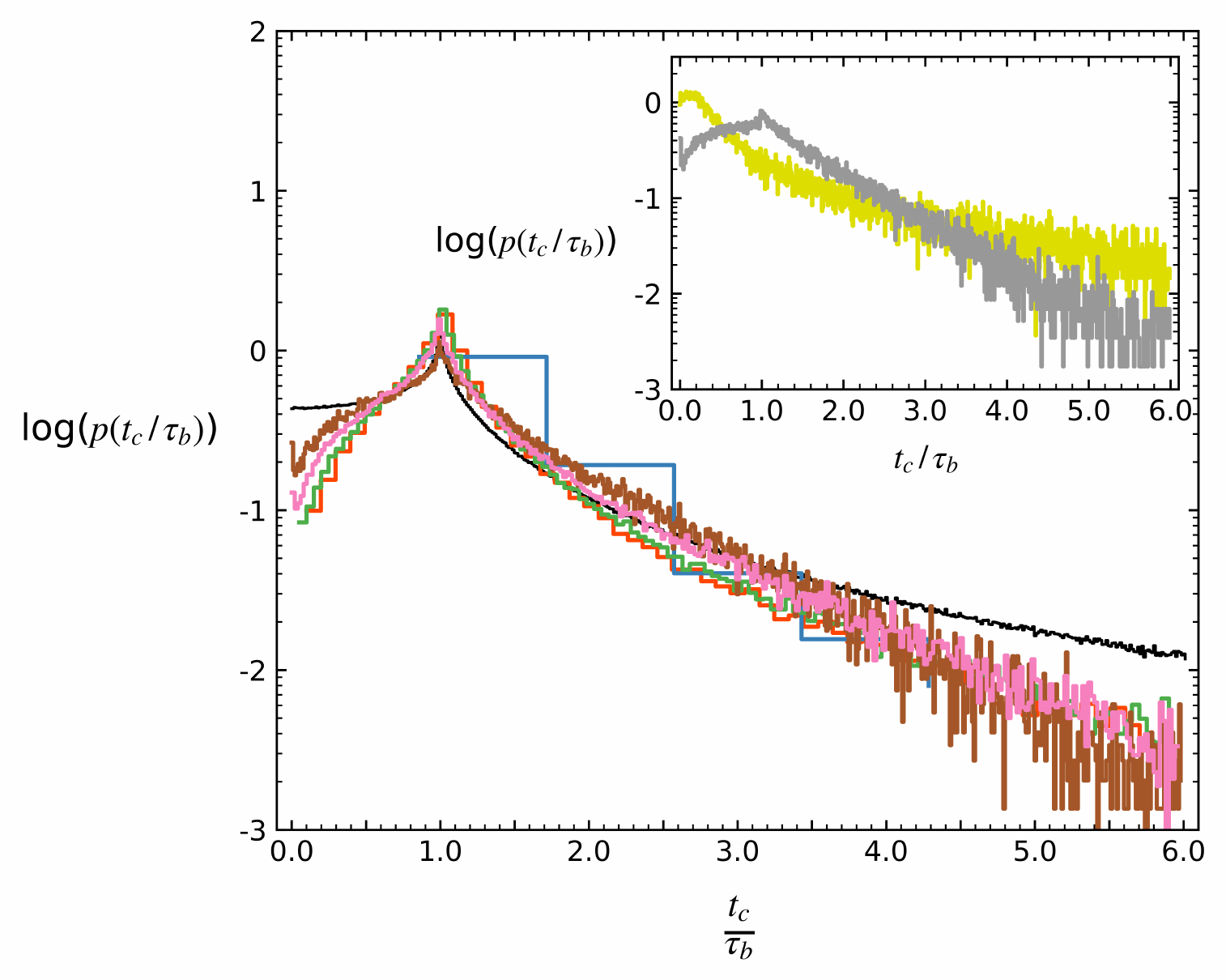}}
\caption{
Contact duration $tc$ distributions for simulations differing in particle size only, scaled by the ballistic crossing time $\tau_b={{2a}/{\delta r}}$.  Colors indicate distributions obtained from simulations with walker radii $a\approx[0.004, 0.04, 0.08, 0.2, 0.4]\,\delta r$ or equivalently about $[0.00071, 0.0071, 0.014, 0.035, 0.071]$ times the mean nearest-neighbor distance between walkers
(the same simulations and color scheme as Figure~\ref{fig7} and~\ref{fig8}).  Contact duration distribution for randomly oriented ballistic intersections is indicated with the underlying {\it solid black} curve.  Inset shows waiting-time distributions for simulations with  $a\approx[0.6, 2.8]\,\delta r$ or about $[0.1, 0.5]$ times the mean nearest-neighbor distance ({\it grey} and {\it yellow} respectively).  
Note that, for the smallest walker radius ({\it blue}), the numerical sub-step employed is insufficient to fully resolve the $t_c$ distribution over the narrow range around  $\tau_b$ or capture contacts shorter than the head-on ballistic crossing time.}
\label{fig9}
\end{figure}

Under this ballistic-crossing approximation, the contact-duration distribution is given by Equation~\ref{eqn12}, with $\theta_1$ and $\theta_2$ independently and uniformly distributed between zero and $2\pi$.  It is plotted, for the values of $2a$ and $\delta r$ used in the simulations,  in the lower panel of Figure~\ref{fig8}.  Each distribution peaks at $\tau_b=2a/\delta r$, the ballistic-contact time for head-on collisions (marked with vertical {\it dashed} fiducial lines).  The distribution wing to shorter contact durations reflects the intersection-chord-length distribution for walkers moving in opposite, but not head-on, directions, and wing to longer contact durations reflects the extended contact periods that can occur between walkers traveling in the same direction, between walkers with low relative velocities.  The idealize ballistic distributions are self-similar, overlapping when $t_c$ is scaled by $\tau_b$; 
the distributions in the lower panel of Figure~\ref{fig8} are identical when the contact duration is rescaled by the ballistic crossing time.

The contact-duration distributions measured in the simulations (upper panel of Figure~\ref{fig8}) show similar features, also peaking at $\tau_b=2a/\delta r$ but with differently shaped wings and some sensitivity to the walker radii (the step size and number density are constant over these runs).  In the simulations there is a finite probability of directional change by one of the walkers during contact. This non-ballistic contribution to the motion flattens the distribution peak and lessen the probability of very short or very long duration contacts (in comparison to the idealized purely-ballistic distributions).  With increased particle radius, the probability of directional change during contact increases.  For very large particle radii, the interaction distance approaches the step size ({\it grey} curve, $2a\approx1.2\,\delta r$) and the mean nearest-neighbor separation ({\it yellow} curve, $2a\approx0.5/\sqrt{n}$), and the distributions lose the ballistic-crossing peak altogether.  This occurs in the first case, because a change in the direction of particle motions occurs during most every encounter, and in the second because multi-particle contact becomes more frequent, blending the contact durations.   

Rescaling $t_c$ by $\tau_b$ highlights these non-ballistic effects and the residual sensitivity of the contact-duration distributions to particle radius. Figure~\ref{fig9} displays the distributions from Figure~\ref{fig8} after rescaling $t_c$ by $\tau_b$.  The underlying {\it black} curve plots the randomly-oriented ballistic crossing time distribution.  The simulation distributions largely overlap under this scaling, but show a residual systematic decrease in the amplitude of the peak relative to the wings
with increase in particle size (Figure~\ref{fig9} main body, {\it orange} to {\it brown}, lower amplitude wings to higher).  This trend continues until non-ballistic effects dominant when the walker size becomes comparable to the step-size or to the mean nearest-neighbor distance (Figure~\ref{fig9} inset).   Note that, for the smallest walker radius (distribution shown in {\it blue}), the numerical sub-step employed is insufficient to resolve the $t_c$ distribution over the narrow interval around $\tau_b$ displayed or to capture any contacts shorter than the head-on ballistic crossing time. 

\section{\label{sec10}Conclusion}

In this paper we examined waiting-time, count, and contact-duration statistics in populations of noninteracting random walkers as functions of the interaction distance (particle size), the random walk step length (correlation length of the motions), and the mean separation of the walkers (number density) in the populations.   A particular ordering was chosen for the magnitudes of these parameters, typically $a < \delta r < 0.5/\sqrt{n}$ in the simulations.  In more complex and realistic settings, the length-scale values and their ordering depends on the dynamics of the motions and the physical properties of the population and its members.  

Using random walk simulations, we uncovered non-exponential waiting-time (non-Poission count) behavior associated with a negative-duration dependency of the waiting time interval.  The non-ballistic motions of walkers in close proximity shortens the waiting time to repeat contact.  Since the mean and variance of the separation between two walkers increases with time, the probability of repeat contact between two walkers increases with decreased distance between them, peaking immediately after a previous contact. This increases the probability of the next contact occurring soon after a previous one, leading to the enhancement of short waiting times and over-dispersion of contact counts.  The random directional changes in the motions also modifies the contact duration distribution, which would otherwise reflect strictly ballistic-crossing between individuals.  

Further, we demonstrated that the differences between the waiting-time, contact-count, and contact-duration distributions in different populations are determined by two key time-scales: the mean-free-collision time and ballistic-crossing time.  
Temporal scalings based on these collapse the waiting time and contact count distributions into common forms across different populations, with very small residual differences reflecting particle size sensitivities at close range, and larger deviations occurring when the walker radius (interaction distance) approaches the step size or the mean nearest-neighbor separation in the population.
Similarly, the contact-duration distributions for populations differing in particle size overlap when scaled by the ballistic-crossing time, with the individual distribution shapes again showing small residual sensitivity to finite-particle-size effects at very close range. 

The canonical random-walk induced by elastic collisions between ballistic trajectories, in an ideal gas for example, is a special limiting case of the random-walks we considered here.  It displays strictly Poisson statistics because directional changes are caused by the collisional interaction between the particles themselves.  In Brownian motion this is not the case.  The random walk characteristics are determined separately from the contacts between the individuals undergoing the Brownian motion.  Examples in natural systems range from biological, in which the random walk characteristics may be determined by behavior, to physical, in which the motion of a dilute component may be governed by collisions between it and the primary component or by an underlying turbulent flow.
This paper focused on the simplest case in which each individual in the population undertakes an independent random walk in two dimensions.
The results are most relevant to systems in which particle proximity is critical and contact results in no change (or a low probability of change) in the particles' motions, or to systems in which multiple contact encounters are required before interaction.  
They also form the basis for 
follow-on work which will look at how the population contact statistics reported here change with more complex underlying dynamics.  
This includes more complex flow dynamics, such as turbulent flows which show non-diffusive Lagrangian transport~\citep[e.g.,][and references therein]{2016PhRvE..93d3120R} and more complex models of particle contact, including particle interaction and non-overlap (volume exclusion).  The effect of particle interaction has been previously evaluated for many particle diffusive systems using macroscopic fluctuation theory~\citep{2015RvMP...87..593B} in the context of both occupation times in one dimension~\citep{2019PhRvE..99e2102A} and the short and long time limits of the non-escape probability from a bounded domain~\citep{2018PhRvL.120l0601A}.  The importance of particle volume exclusion has been recently assessed using the boundary local time distribution~\citep[e.g.,][and references therein]{2019PhRvE.100f2110G} in the context of both first-passage time problems~\citep{2020JSMTE2020j3205G} and contact counts between two Brownian particles on a plane~\citep{2021JPhA...54a5003G}. 

One important application of the work presented in this paper may be contagion in human populations.    
The motions of individuals in populations may be largely determined independently from contact events, as in our model, though elements of collective behavior~\citep[e.g.,][]{doi:10.1177/0963721417746743} may also be present.  
Individuals cross each others' paths within an interaction distance (contagion radius) of each other, the overlap of contagion zones does not necessarily cause trajectory changes, and the interaction distance is often smaller than the correlation length of the motions.  Under these circumstances, contacts are unlikely to show exponential waiting-time distributions and corresponding Poisson counts, and the contact-duration distribution may be peaked around the ballistic-crossing time of two individuals.
As the number density of individuals varies between populations, the time scale over which non-exponential contact statistics are apparent should as well.  It would be interesting to test the limits of these suggestions using high resolution cell-phone location data~\citep{zhao2021}.       

\vskip 0.25in
\noindent {\bf Acknowledgements:}  
Special thanks to 

\appendix

\section{\label{appendix0}Pair separation in two dimensions}

Underlying the contact statistics in populations of random walkers are the statistics of pair separation, as the waiting time interval between contacts (the inter-arrival time) for any individual is the minimum statistic, over all other individuals, of the first passage time to a specified contact-distance given the pair-separation distribution of the population. 
While, in the Brownian limit, the diffusion equation readily yields a closed form solution for the random walk first-passage time distribution to a point or sphere in one or three dimensions (the Levy distribution), it fails to yield such a convenient solution in two-dimensions.
The fundamental underlying difficulty arises because, while the motion of each individual on a two dimensional plane can be described as a two-dimensional random walk, the distance between any two walkers only approximates a one-dimensional constant diffusivity process in the short and long time limits.  At intermediate times the pair-separation probability distribution evolves from approximately Gaussian to Rayleigh with a corresponding nonlinear change in the variance.     

In two dimensions, the distance $r_s$ between two unbiased random walkers, as a function of their initial separation $r_0$ at $t=t_0$, is Rice distributed
\begin{equation}
p(r_s\vert r_0)  = {{r_s}\over{\sigma^2}}\exp\left(-{{r_s^2+r_0^2}\over{2\sigma^2}}\right)I_0\left({{r_0r_s}\over{\sigma^2}}\right)\ ,
\label{eqn2b}
\end{equation}
with $I_0$ denoting the lowest order modified Bessel function of the first kind~\cite{1972NBSAbramowitzandStegun} and the scale parameter 
$\sigma^2= 4\,D\,(t-t_0)$, with $D=\delta r^2 /4 \delta t$, in the continuous time and space limit.  
The pair-separation variance (the variance of the Rice distribution, Equation~\ref{eqn2b}) is a nonlinear function of the scale parameter $\sigma^2= 4\,D\,(t-t_0)$, and thus time,
\begin{equation}
\label{eqn3}
\sigma_{\! s}^2=2\sigma^2+r_0^2-{{\pi\sigma^2}\over{2}} L_{1/2}^2\left(-r_0^2/2\sigma^2\right)\ ,  
\end{equation}
where $L_{1/2}^2$ indicates the square of the $L_{1/2}$ Laguerre polynomial.    

\begin{figure}[t!]
\centerline{\includegraphics[width=8.0cm,trim=0.0cm 0.0cm 1.0cm 0.75cm]{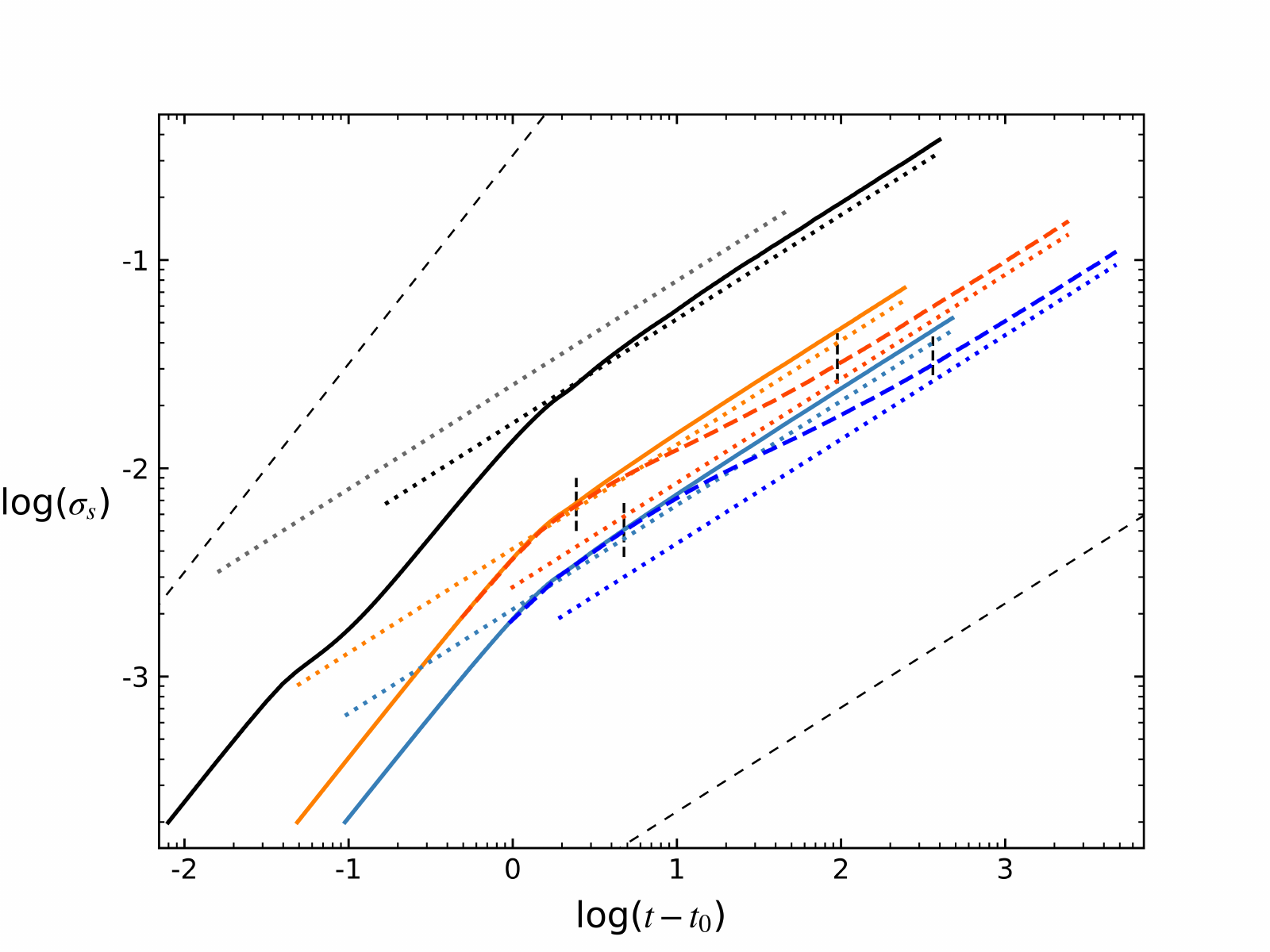}}
\caption{Standard deviation of the pair separation distribution as a function of time (in units of the number of steps taken) for random walkers of varying initial separation and step length.  Curves of {\it solid} colors indicate pair separation results for walks with step lengths ({\it left} to {\it right}) $\delta r =[0.0041, 0.0021]$ and initial separation of $r_0=0.7$.  {\it Black} curve indicates standard deviation of the separation for a pair with step length $\delta r = 0.0251$ and initial separation $r_0=0.002$.  The latter are values typical of those immediately after contact between pairs in the walker populations simulations discussed in Section~\ref{sec4.1}.  {\it Dotted} curves of the same color indicate the expected standard deviations in the diffusion limit as discussed in the text.  Dark {\it orange} and {\it blue dashed} curves plot the pair separation distribution standard deviation for two pairs of random walkers with step length $\delta r = [0.0041, 0.0021]$ and initial separation $r_0=0.02$. Small {\it dotted} vertical fiducial lines indicate the time of first possible contact between those pairs and the approximate time of transition to the long-time Rayleigh distribution (see Figure~\ref{fig2} and discussion in text).}
\label{fig1}
\end{figure}

The standard deviation of the separation between pairs in random-walk simulations is plotted as a function of time in Figure~\ref{fig1}.  Time is measured in units of the number of steps taken (i.e., $\delta t=1$).  The simulations follow those described in Section~\ref{sec2} of the main text for individual pairs rather than populations, but are of short enough duration that periodicity plays no role. 
We use them to measure pair separation as a function of time for different values of the initial separation $r_0$ 
and mean step length $\delta r$.  
The {\it solid} color (non-black) curves in Figure~\ref{fig1} plot $\sigma_{\! s}$ for initially well separated ($r_0=0.7$) pairs taking steps of mean length 
$\delta r =0.0041$ ({\it solid orange} curve) and $\delta r =0.0021$ ({\it solid blue} curve).    
As expected, the distance between two random walkers scales ballistically, $\sigma_s\sim t$, for times short compared to one
step and diffusively,  $\sigma_{\! s}\sim t^{1/2}$, for times greater than this. 
For reference the $\sigma_{\! s}\sim t$ and $\sigma_{\! s}\sim t^{1/2}$ scalings are indicated with {\it black dashed} lines, and the strict diffusive limit for each individual simulation, $\sigma_{\! s}=\sigma= \sqrt{4\,D\,(t-t_0)}$, is plotted as a {\it dotted} curve of the same color. The offset between the theoretical diffusive limit and numerically determined result at long times (the offset between the {\it dotted} lines and and {\it solid} curves of the same color at long times) reflects the discrete early ballistic motion, which is resolved numerically for times less than one by taking small sub-steps (see Section~\ref{sec2} of main text).

\begin{figure}[t!]
\centerline{\includegraphics[width=8.0cm,trim=0.0cm 1.0cm 1.0cm 1.0cm]{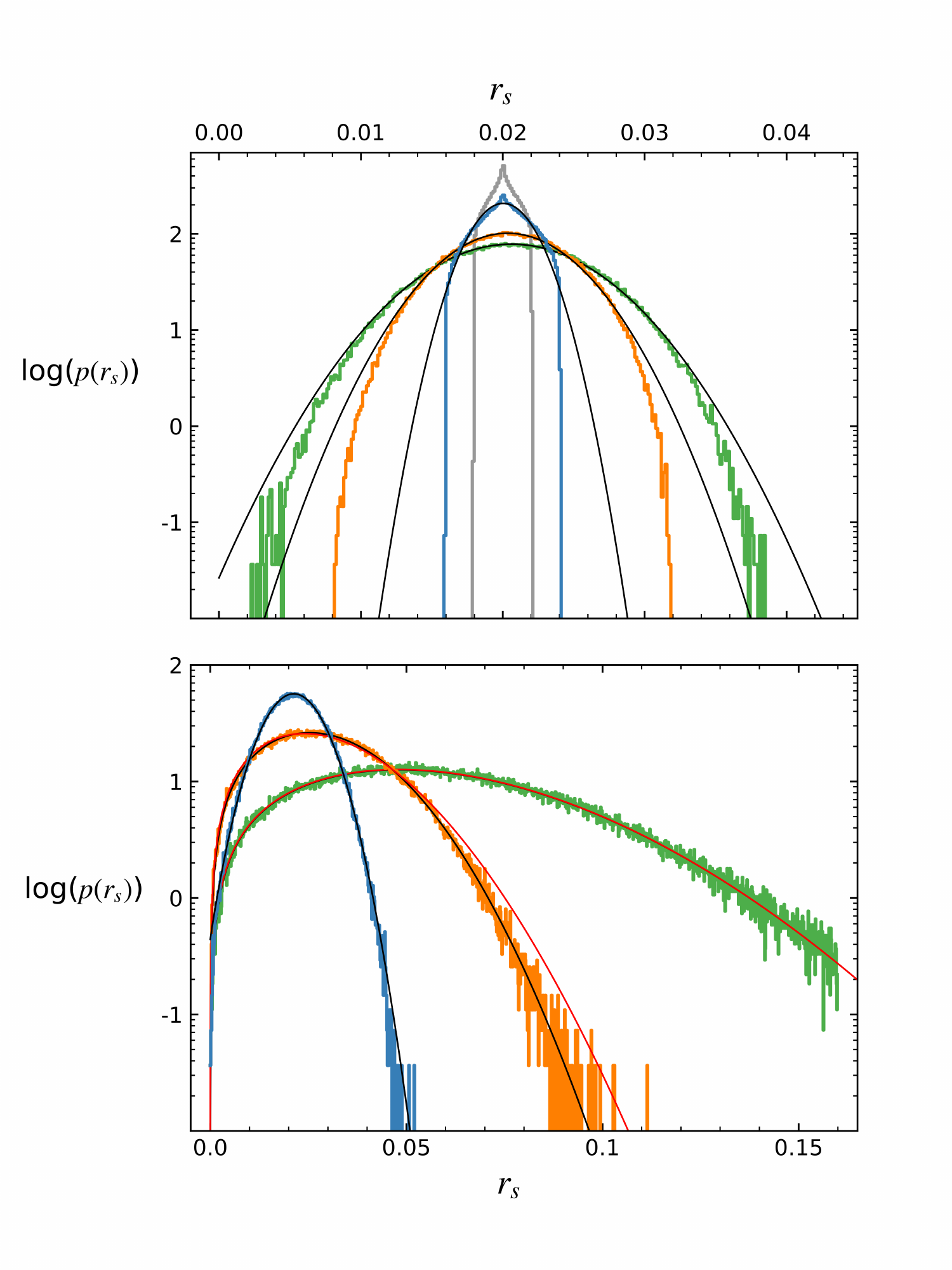}}
\caption{Snapshots of the pair-separation probability density as a function of time for pairs of walkers whose separation variance is plotted with the {\it dashed dark blue} curve in Figure~\ref{fig1}.  Distributions are shown for $t-t_0=[0.5,1.0, 3.0, r_0/(2\delta r)=4.76]$ (narrow to wider pdfs respectively) in {\it upper} panel, and for $t-t_0=[10.0, 60.0,  4 r_0^2/\delta r^2=363]$ (narrow to wider pdfs respectively) in {\it lower} panel.  In {\it upper} panel, {\it black} curves indicate best fit Gaussian distributions, and in the {\it lower} panel {\it black} curves  indicate best fit Rice distributions.  Best fit Rayleigh distributions are shown in {\it red} in the {\it lower} panel.}
\label{fig2}
\end{figure}

The same simulations were also run for smaller initial pair separation ($r_0=0.02$).  
The resulting evolution of the pair-separation standard deviation is plotted in Figure~\ref{fig1} using {\it dashed} line styles. 
They illustrates how the {\it solid} curves would behave if extended to longer times.  
The curves deviate from diffusive scaling at $t-t_0\approx r_0/(2\delta r)$ 
(marked by the lefthand short {\it dashed} vertical fiducial lines for each case).  
This is the earliest possible time (time measured in mean step duration, ${{\delta t}}=1$) that the distance between the two random walkers can equal zero.  Diffusive scaling is re-established later, but with a reduced variance equal to $\sigma_s^2=(4-\pi)/2\ \sigma^2$ (indicated with {\it dottted dark orange} and {\it dotted dark blue} liness).  
By Equation~\ref{eqn3}, this occurs for $r_0^2\ll 2\sigma^2$, and the approximate time at which it occurs $t-t_0\approx4 r_0^2/\delta r^2$ is marked by the two righthand short {\it dashed} vertical fiducial lines.

The change in variance between $t-t_0\approx r_0/(2\delta r)$ and $t-t_0\approx4 r_0^2/\delta r^2$ reflects a change in the underlying separation probability density function.  In two dimensions, this depends critically on both the initial separation and the elapsed time.   
Figure~\ref{fig2} shows the temporal evolution of the normalized pair separation distribution for pairs of walkers whose separation variance is plotted with the {\it dashed dark blue} curve in Figure~\ref{fig1}.  For times shorter than one step, the probability density function is nonGaussian ({\it grey} curve in top panel of Figure~\ref{fig2}), reflecting the ballistic separation of the walkers (Equation~\ref{eqn4}, main text).
This ballistic phase is important in determining the contact duration distributions (as discussed Section~\ref{sec5} of the main text).  
For somewhat longer but still short-times, $1.0\lesssim t-t_0\leq r_0/(2\delta r)$ (illustrated by the remaining, non-{\it grey} curves in {\it upper} panel of Figure~\ref{fig2}), the pair separation is distributed as a truncated Gaussian, with the truncation occurring at a value of $r=r_0\pm2\,\delta r\,(t-t_0)$, the maximum and minimum separations that the walkers can achieve over the 
elapsed time.  
During this period the separation distribution variance scales diffusively, with a diffusivity equal to twice that of the displacement of an individual walker from its origin, $\sigma_s^2= \sigma^2={4\,D\,(t-t_0)}$. For times longer than $t-t_0 = r_0/(2\delta r)$ ({\it lower} panel), the pair separation is Rice distributed, with the mean shifting to larger separations with time and the variance growing more slowly than $t$ because separation values reflect across $r=0$.  At long times, $t-t_0\gtrsim4 r_0^2/\delta r^2$, the Rice distribution asymptotes to Rayleigh, with the variance again scaling with $t-t_0$ but offset to a reduced value $\sigma_s^2=(4-\pi)/2\ \sigma^2$.   These behaviors, while making direct calculation of the first passage time difficult, provide an opportunity for simplified analysis in the short and long time limits (Section~\ref{sec4.2}, main text).

It is important to note that, for initial separations smaller than the step size, the shift to Rayleigh distributed walker separation occurs during the ballistic phase of the motion (over times shorter than about one step).
This is illustrated by {\it black} curve in Figure~\ref{fig1}).  In that simulation, the initial separation between walkers was taken to be twice a typical walker radius $a$ employed in the many-walker population simulations described in the main text ($r_0=2a=0.002$).  
This is the separation walkers would have immediately after contact in those runs.  
The step size was take to be a factor of about 12.5 greater than this ($\delta r=0.0251$), again a value typical of most of the simulations undertaken for the main body of this paper.  With these parameters, the pair separation distribution becomes Rayleigh before the variance scaling transitions from ballistic to diffusive, and the variance asymptotes directly to its Rayleigh distribution value.  
We note that, depending on the ratio of the particle to step sizes, the transition to Rayleigh-distributed walker separation and diffusive scaling after contact can occur at any time with respect to the ballistic to diffusive scaling transition. 

For most of the population simulations discussed in the main text, the walker interaction distance was taken to be much smaller than the step length, and thus, for simulation times longer than about one step after each contact, the separation between any two walkers is already Rayleigh distributed with the mean separation increasing as the square-root of time and the variance increasing linearly with time.  This is consistent with the approximate solution derived in Section~\ref{sec4.2} of the main text. 

\bibliography{randomwalk}

\begin{thebibliography}{61}
\expandafter\ifx\csname natexlab\endcsname\relax\def\natexlab#1{#1}\fi
\expandafter\ifx\csname bibnamefont\endcsname\relax
  \def\bibnamefont#1{#1}\fi
\expandafter\ifx\csname bibfnamefont\endcsname\relax
  \def\bibfnamefont#1{#1}\fi
\expandafter\ifx\csname citenamefont\endcsname\relax
  \def\citenamefont#1{#1}\fi
\expandafter\ifx\csname url\endcsname\relax
  \def\url#1{\texttt{#1}}\fi
\expandafter\ifx\csname urlprefix\endcsname\relax\def\urlprefix{URL }\fi
\providecommand{\bibinfo}[2]{#2}
\providecommand{\eprint}[2][]{\url{#2}}

\bibitem[{\citenamefont{{Redner}}(2007)}]{2007gfpp.book.....R}
\bibinfo{author}{\bibfnamefont{S.}~\bibnamefont{{Redner}}},
  \emph{\bibinfo{title}{{A Guide to First-Passage Processes}}}
  (\bibinfo{publisher}{Cambridge University Press, Cambridge},
  \bibinfo{year}{2007}).

\bibitem[{\citenamefont{{Hu} et~al.}(2010)\citenamefont{{Hu}, {Cheng}, and
  {Berne}}}]{2010JChPh.133c4105H}
\bibinfo{author}{\bibfnamefont{Z.}~\bibnamefont{{Hu}}},
  \bibinfo{author}{\bibfnamefont{L.}~\bibnamefont{{Cheng}}}, \bibnamefont{and}
  \bibinfo{author}{\bibfnamefont{B.~J.} \bibnamefont{{Berne}}},
  \bibinfo{journal}{\jcp} \textbf{\bibinfo{volume}{133}},
  \bibinfo{pages}{034105} (\bibinfo{year}{2010}).

\bibitem[{\citenamefont{{Chandrasekhar}}(1943)}]{1943RvMP...15....1C}
\bibinfo{author}{\bibfnamefont{S.}~\bibnamefont{{Chandrasekhar}}},
  \bibinfo{journal}{Reviews of Modern Physics} \textbf{\bibinfo{volume}{15}},
  \bibinfo{pages}{1} (\bibinfo{year}{1943}).

\bibitem[{\citenamefont{Nelson}(1967)}]{nelson1967dynamical}
\bibinfo{author}{\bibfnamefont{E.}~\bibnamefont{Nelson}},
  \emph{\bibinfo{title}{Dynamical Theories of Brownian Motion}}, Mathematical
  Notes - Princeton University Press (\bibinfo{publisher}{Princeton University
  Press}, \bibinfo{year}{1967}).

\bibitem[{\citenamefont{Weiss}(1994)}]{weiss1994aspects}
\bibinfo{author}{\bibfnamefont{G.}~\bibnamefont{Weiss}},
  \emph{\bibinfo{title}{Aspects and Applications of the Random Walk}},
  International Congress Series (\bibinfo{publisher}{North-Holland},
  \bibinfo{year}{1994}).

\bibitem[{\citenamefont{Peres et~al.}(2010)\citenamefont{Peres, Schramm,
  Werner, and M{\"o}rters}}]{peres2010brownian}
\bibinfo{author}{\bibfnamefont{Y.}~\bibnamefont{Peres}},
  \bibinfo{author}{\bibfnamefont{O.}~\bibnamefont{Schramm}},
  \bibinfo{author}{\bibfnamefont{W.}~\bibnamefont{Werner}}, \bibnamefont{and}
  \bibinfo{author}{\bibfnamefont{P.}~\bibnamefont{M{\"o}rters}},
  \emph{\bibinfo{title}{Brownian Motion}}, Cambridge Series in Statistical and
  Probabilistic Mathematics (\bibinfo{publisher}{Cambridge University Press},
  \bibinfo{year}{2010}).

\bibitem[{\citenamefont{Metzler et~al.}(2014)\citenamefont{Metzler, Redner, and
  Oshanin}}]{metzler2014first}
\bibinfo{author}{\bibfnamefont{R.}~\bibnamefont{Metzler}},
  \bibinfo{author}{\bibfnamefont{S.}~\bibnamefont{Redner}}, \bibnamefont{and}
  \bibinfo{author}{\bibfnamefont{G.}~\bibnamefont{Oshanin}},
  \emph{\bibinfo{title}{First-passage Phenomena And Their Applications}}, World
  Scientific Studies in International Economics (\bibinfo{publisher}{World
  Scientific Publishing Company}, \bibinfo{year}{2014}).

\bibitem[{\citenamefont{Libchaber}(2019)}]{annurev-conmatphys2019}
\bibinfo{author}{\bibfnamefont{A.}~\bibnamefont{Libchaber}},
  \bibinfo{journal}{Annual Review of Condensed Matter Physics}
  \textbf{\bibinfo{volume}{10}}, \bibinfo{pages}{275} (\bibinfo{year}{2019}).

\bibitem[{\citenamefont{Grebenkov et~al.}(2020)\citenamefont{Grebenkov,
  Holcman, and Metzler}}]{Grebenkov_2020}
\bibinfo{author}{\bibfnamefont{D.~S.} \bibnamefont{Grebenkov}},
  \bibinfo{author}{\bibfnamefont{D.}~\bibnamefont{Holcman}}, \bibnamefont{and}
  \bibinfo{author}{\bibfnamefont{R.}~\bibnamefont{Metzler}},
  \bibinfo{journal}{Journal of Physics A: Mathematical and Theoretical}
  \textbf{\bibinfo{volume}{53}}, \bibinfo{pages}{190301}
  (\bibinfo{year}{2020}).

\bibitem[{\citenamefont{{Sanders}}(2009)}]{2009PhRvE..80c6119S}
\bibinfo{author}{\bibfnamefont{D.~P.} \bibnamefont{{Sanders}}},
  \bibinfo{journal}{\pre} \textbf{\bibinfo{volume}{80}}, \bibinfo{eid}{036119}
  (\bibinfo{year}{2009}).

\bibitem[{\citenamefont{{Zhang} et~al.}(2011)\citenamefont{{Zhang}, {Julaiti},
  {Hou}, {Zhang}, and {Chen}}}]{2011EPJB...84..691Z}
\bibinfo{author}{\bibfnamefont{Z.}~\bibnamefont{{Zhang}}},
  \bibinfo{author}{\bibfnamefont{A.}~\bibnamefont{{Julaiti}}},
  \bibinfo{author}{\bibfnamefont{B.}~\bibnamefont{{Hou}}},
  \bibinfo{author}{\bibfnamefont{H.}~\bibnamefont{{Zhang}}}, \bibnamefont{and}
  \bibinfo{author}{\bibfnamefont{G.}~\bibnamefont{{Chen}}},
  \bibinfo{journal}{European Physical Journal B} \textbf{\bibinfo{volume}{84}},
  \bibinfo{pages}{691} (\bibinfo{year}{2011}).

\bibitem[{\citenamefont{{Riascos} and {Sanders}}(2021)}]{2021PhRvE.103d2312R}
\bibinfo{author}{\bibfnamefont{A.~P.} \bibnamefont{{Riascos}}}
  \bibnamefont{and} \bibinfo{author}{\bibfnamefont{D.~P.}
  \bibnamefont{{Sanders}}}, \bibinfo{journal}{\pre}
  \textbf{\bibinfo{volume}{103}}, \bibinfo{eid}{042312} (\bibinfo{year}{2021}).

\bibitem[{\citenamefont{{Tejedor} et~al.}(2011)\citenamefont{{Tejedor},
  {Schad}, {B{\'e}nichou}, {Voituriez}, and {Metzler}}}]{2011JPhA...44M5005T}
\bibinfo{author}{\bibfnamefont{V.}~\bibnamefont{{Tejedor}}},
  \bibinfo{author}{\bibfnamefont{M.}~\bibnamefont{{Schad}}},
  \bibinfo{author}{\bibfnamefont{O.}~\bibnamefont{{B{\'e}nichou}}},
  \bibinfo{author}{\bibfnamefont{R.}~\bibnamefont{{Voituriez}}},
  \bibnamefont{and}
  \bibinfo{author}{\bibfnamefont{R.}~\bibnamefont{{Metzler}}},
  \bibinfo{journal}{Journal of Physics A: Mathematical and Theoretical}
  \textbf{\bibinfo{volume}{44}}, \bibinfo{eid}{395005} (\bibinfo{year}{2011}).

\bibitem[{\citenamefont{{Le Vot} et~al.}(2020)\citenamefont{{Le Vot}, {Yuste},
  {Abad}, and {Grebenkov}}}]{2020PhRvE.102c2118L}
\bibinfo{author}{\bibfnamefont{F.}~\bibnamefont{{Le Vot}}},
  \bibinfo{author}{\bibfnamefont{S.~B.} \bibnamefont{{Yuste}}},
  \bibinfo{author}{\bibfnamefont{E.}~\bibnamefont{{Abad}}}, \bibnamefont{and}
  \bibinfo{author}{\bibfnamefont{D.~S.} \bibnamefont{{Grebenkov}}},
  \bibinfo{journal}{\pre} \textbf{\bibinfo{volume}{102}}, \bibinfo{eid}{032118}
  (\bibinfo{year}{2020}).

\bibitem[{\citenamefont{Lavine}(2005)}]{2005Lavine}
\bibinfo{author}{\bibfnamefont{J.}~\bibnamefont{Lavine}}, \bibinfo{journal}{MRS
  Online Proceedings Library} \textbf{\bibinfo{volume}{899}},
  \bibinfo{pages}{713} (\bibinfo{year}{2005}).

\bibitem[{\citenamefont{Enns et~al.}(1984)\citenamefont{Enns, Smith, and
  Ehlers}}]{10.2307/3213665}
\bibinfo{author}{\bibfnamefont{E.~G.} \bibnamefont{Enns}},
  \bibinfo{author}{\bibfnamefont{B.~R.} \bibnamefont{Smith}}, \bibnamefont{and}
  \bibinfo{author}{\bibfnamefont{P.~F.} \bibnamefont{Ehlers}},
  \bibinfo{journal}{Journal of Applied Probability}
  \textbf{\bibinfo{volume}{21}}, \bibinfo{pages}{70} (\bibinfo{year}{1984}).

\bibitem[{\citenamefont{{Basnayake} et~al.}(2019)\citenamefont{{Basnayake},
  {Schuss}, and {Holcman}}}]{2019JNS....29..461B}
\bibinfo{author}{\bibfnamefont{K.}~\bibnamefont{{Basnayake}}},
  \bibinfo{author}{\bibfnamefont{Z.}~\bibnamefont{{Schuss}}}, \bibnamefont{and}
  \bibinfo{author}{\bibfnamefont{D.}~\bibnamefont{{Holcman}}},
  \bibinfo{journal}{Journal of NonLinear Science}
  \textbf{\bibinfo{volume}{29}}, \bibinfo{pages}{461} (\bibinfo{year}{2019}).

\bibitem[{\citenamefont{{Lawley} and {Miles}}(2019)}]{2019JNS....29.2955L}
\bibinfo{author}{\bibfnamefont{S.~D.} \bibnamefont{{Lawley}}} \bibnamefont{and}
  \bibinfo{author}{\bibfnamefont{C.~E.} \bibnamefont{{Miles}}},
  \bibinfo{journal}{Journal of NonLinear Science}
  \textbf{\bibinfo{volume}{29}}, \bibinfo{pages}{2955} (\bibinfo{year}{2019}).

\bibitem[{\citenamefont{{Lawley}}(2020{\natexlab{a}})}]{2020PhRvE.101a2413L}
\bibinfo{author}{\bibfnamefont{S.~D.} \bibnamefont{{Lawley}}},
  \bibinfo{journal}{\pre} \textbf{\bibinfo{volume}{101}}, \bibinfo{eid}{012413}
  (\bibinfo{year}{2020}{\natexlab{a}}).

\bibitem[{\citenamefont{{Nayak} et~al.}(2020)\citenamefont{{Nayak}, {Nandi},
  and {Das}}}]{2020PhRvE.102f2109N}
\bibinfo{author}{\bibfnamefont{I.}~\bibnamefont{{Nayak}}},
  \bibinfo{author}{\bibfnamefont{A.}~\bibnamefont{{Nandi}}}, \bibnamefont{and}
  \bibinfo{author}{\bibfnamefont{D.}~\bibnamefont{{Das}}},
  \bibinfo{journal}{\pre} \textbf{\bibinfo{volume}{102}}, \bibinfo{eid}{062109}
  (\bibinfo{year}{2020}).

\bibitem[{\citenamefont{{Changruenngam}
  et~al.}(2020)\citenamefont{{Changruenngam}, {Bicout}, and
  {Modchang}}}]{Changruenngam2020}
\bibinfo{author}{\bibfnamefont{S.}~\bibnamefont{{Changruenngam}}},
  \bibinfo{author}{\bibfnamefont{D.}~\bibnamefont{{Bicout}}}, \bibnamefont{and}
  \bibinfo{author}{\bibfnamefont{C.}~\bibnamefont{{Modchang}}},
  \bibinfo{journal}{Scientific Reports} \textbf{\bibinfo{volume}{10}},
  \bibinfo{pages}{11325} (\bibinfo{year}{2020}).

\bibitem[{\citenamefont{Norambuena et~al.}(2020)\citenamefont{Norambuena,
  Valencia, and Guzm{\'a}an-Lastra}}]{norambuena2020}
\bibinfo{author}{\bibfnamefont{A.}~\bibnamefont{Norambuena}},
  \bibinfo{author}{\bibfnamefont{F.}~\bibnamefont{Valencia}}, \bibnamefont{and}
  \bibinfo{author}{\bibfnamefont{F.}~\bibnamefont{Guzm{\'a}an-Lastra}},
  \bibinfo{journal}{Scientific Reports} \textbf{\bibinfo{volume}{10}},
  \bibinfo{pages}{20845} (\bibinfo{year}{2020}).

\bibitem[{\citenamefont{{Verma} et~al.}(2020)\citenamefont{{Verma},
  {Bhatnagar}, {Mitra}, and {Pandit}}}]{2020PhRvR...2c3239V}
\bibinfo{author}{\bibfnamefont{A.~K.} \bibnamefont{{Verma}}},
  \bibinfo{author}{\bibfnamefont{A.}~\bibnamefont{{Bhatnagar}}},
  \bibinfo{author}{\bibfnamefont{D.}~\bibnamefont{{Mitra}}}, \bibnamefont{and}
  \bibinfo{author}{\bibfnamefont{R.}~\bibnamefont{{Pandit}}},
  \bibinfo{journal}{Physical Review Research} \textbf{\bibinfo{volume}{2}},
  \bibinfo{eid}{033239} (\bibinfo{year}{2020}).

\bibitem[{\citenamefont{{Andrews} and {Bray}}(2004)}]{2004PhBio...1..137A}
\bibinfo{author}{\bibfnamefont{S.~S.} \bibnamefont{{Andrews}}}
  \bibnamefont{and} \bibinfo{author}{\bibfnamefont{D.}~\bibnamefont{{Bray}}},
  \bibinfo{journal}{Physical Biology} \textbf{\bibinfo{volume}{1}},
  \bibinfo{pages}{137} (\bibinfo{year}{2004}).

\bibitem[{\citenamefont{{Grebenkov}}(2019{\natexlab{a}})}]{grebenkov2019}
\bibinfo{author}{\bibfnamefont{D.~S.} \bibnamefont{{Grebenkov}}}, in
  \emph{\bibinfo{booktitle}{Chemical Kinetics: Beyond the Textbook}}, edited by
  \bibinfo{editor}{\bibfnamefont{K.}~\bibnamefont{{Lindenberg}}},
  \bibinfo{editor}{\bibfnamefont{R.}~\bibnamefont{{Metzler}}},
  \bibnamefont{and} \bibinfo{editor}{\bibfnamefont{G.}~\bibnamefont{{Oshanin}}}
  (\bibinfo{publisher}{World Scientific Publishing Company},
  \bibinfo{year}{2019}{\natexlab{a}}), pp. \bibinfo{pages}{191--219}.

\bibitem[{\citenamefont{{Grebenkov}}(2020{\natexlab{a}})}]{2020PhRvL.125g8102G}
\bibinfo{author}{\bibfnamefont{D.~S.} \bibnamefont{{Grebenkov}}},
  \bibinfo{journal}{\prl} \textbf{\bibinfo{volume}{125}}, \bibinfo{eid}{078102}
  (\bibinfo{year}{2020}{\natexlab{a}}).

\bibitem[{\citenamefont{{Kempf} et~al.}(1999)\citenamefont{{Kempf}, {Pfalzner},
  and {Henning}}}]{1999Icar..141..388K}
\bibinfo{author}{\bibfnamefont{S.}~\bibnamefont{{Kempf}}},
  \bibinfo{author}{\bibfnamefont{S.}~\bibnamefont{{Pfalzner}}},
  \bibnamefont{and} \bibinfo{author}{\bibfnamefont{T.~K.}
  \bibnamefont{{Henning}}}, \bibinfo{journal}{Icarus}
  \textbf{\bibinfo{volume}{141}}, \bibinfo{pages}{388} (\bibinfo{year}{1999}).

\bibitem[{\citenamefont{{Homann} et~al.}(2016)\citenamefont{{Homann},
  {Guillot}, {Bec}, {Ormel}, {Ida}, and {Tanga}}}]{2016A&A...589A.129H}
\bibinfo{author}{\bibfnamefont{H.}~\bibnamefont{{Homann}}},
  \bibinfo{author}{\bibfnamefont{T.}~\bibnamefont{{Guillot}}},
  \bibinfo{author}{\bibfnamefont{J.}~\bibnamefont{{Bec}}},
  \bibinfo{author}{\bibfnamefont{C.~W.} \bibnamefont{{Ormel}}},
  \bibinfo{author}{\bibfnamefont{S.}~\bibnamefont{{Ida}}}, \bibnamefont{and}
  \bibinfo{author}{\bibfnamefont{P.}~\bibnamefont{{Tanga}}},
  \bibinfo{journal}{\aap} \textbf{\bibinfo{volume}{589}}, \bibinfo{eid}{A129}
  (\bibinfo{year}{2016}).

\bibitem[{\citenamefont{{Mordant} et~al.}(2004)\citenamefont{{Mordant},
  {L{\'e}v{\^e}que}, and {Pinton}}}]{2004NJPh....6..116M}
\bibinfo{author}{\bibfnamefont{N.}~\bibnamefont{{Mordant}}},
  \bibinfo{author}{\bibfnamefont{E.}~\bibnamefont{{L{\'e}v{\^e}que}}},
  \bibnamefont{and} \bibinfo{author}{\bibfnamefont{J.-F.}
  \bibnamefont{{Pinton}}}, \bibinfo{journal}{New Journal of Physics}
  \textbf{\bibinfo{volume}{6}}, \bibinfo{pages}{116} (\bibinfo{year}{2004}).

\bibitem[{\citenamefont{{Rast} and {Pinton}}(2009)}]{2009PhRvE..79d6314R}
\bibinfo{author}{\bibfnamefont{M.~P.} \bibnamefont{{Rast}}} \bibnamefont{and}
  \bibinfo{author}{\bibfnamefont{J.-F.} \bibnamefont{{Pinton}}},
  \bibinfo{journal}{\pre} \textbf{\bibinfo{volume}{79}}, \bibinfo{eid}{046314}
  (\bibinfo{year}{2009}).

\bibitem[{\citenamefont{Polanowski and Sikorski}(2019)}]{polanowski2019}
\bibinfo{author}{\bibfnamefont{P.}~\bibnamefont{Polanowski}} \bibnamefont{and}
  \bibinfo{author}{\bibfnamefont{A.}~\bibnamefont{Sikorski}},
  \bibinfo{journal}{J. Molecular Modeling} \textbf{\bibinfo{volume}{25}}
  (\bibinfo{year}{2019}).

\bibitem[{\citenamefont{Zhao et~al.}(2021)\citenamefont{Zhao, Zeng, and
  Yeung}}]{zhao2021}
\bibinfo{author}{\bibfnamefont{C.}~\bibnamefont{Zhao}},
  \bibinfo{author}{\bibfnamefont{A.}~\bibnamefont{Zeng}}, \bibnamefont{and}
  \bibinfo{author}{\bibfnamefont{C.}~\bibnamefont{Yeung}},
  \bibinfo{journal}{EPJ Data Sci.} \textbf{\bibinfo{volume}{10}},
  \bibinfo{pages}{5} (\bibinfo{year}{2021}).

\bibitem[{\citenamefont{{Donsker}}(1951)}]{Donsker1951}
\bibinfo{author}{\bibfnamefont{M.~D.} \bibnamefont{{Donsker}}},
  \bibinfo{journal}{Mem. Amer. Math.} \textbf{\bibinfo{volume}{6}},
  \bibinfo{pages}{12} (\bibinfo{year}{1951}).

\bibitem[{\citenamefont{Chapman et~al.}(1990)\citenamefont{Chapman, Cowling,
  Burnett, and Cercignani}}]{chapman1990}
\bibinfo{author}{\bibfnamefont{S.}~\bibnamefont{Chapman}},
  \bibinfo{author}{\bibfnamefont{T.}~\bibnamefont{Cowling}},
  \bibinfo{author}{\bibfnamefont{D.}~\bibnamefont{Burnett}}, \bibnamefont{and}
  \bibinfo{author}{\bibfnamefont{C.}~\bibnamefont{Cercignani}},
  \emph{\bibinfo{title}{The Mathematical Theory of Non-uniform Gases: An
  Account of the Kinetic Theory of Viscosity, Thermal Conduction and Diffusion
  in Gases}}, Cambridge Mathematical Library (\bibinfo{publisher}{Cambridge
  University Press}, \bibinfo{year}{1990}).

\bibitem[{\citenamefont{{Paik}}(2014)}]{2014AmJPh..82..602P}
\bibinfo{author}{\bibfnamefont{S.~T.} \bibnamefont{{Paik}}},
  \bibinfo{journal}{American Journal of Physics} \textbf{\bibinfo{volume}{82}},
  \bibinfo{pages}{602} (\bibinfo{year}{2014}).

\bibitem[{\citenamefont{Gumbel}(1962)}]{gumbel1962}
\bibinfo{author}{\bibfnamefont{E.}~\bibnamefont{Gumbel}},
  \emph{\bibinfo{title}{Statistics of Extremes}} (\bibinfo{publisher}{Columbia
  University Press}, \bibinfo{year}{1962}).

\bibitem[{\citenamefont{David and Nagaraja}(2003)}]{davidandnagaraja2003}
\bibinfo{author}{\bibfnamefont{H.~A.} \bibnamefont{David}} \bibnamefont{and}
  \bibinfo{author}{\bibfnamefont{H.~N.} \bibnamefont{Nagaraja}},
  \emph{\bibinfo{title}{Order statistics}} (\bibinfo{publisher}{John Wiley},
  \bibinfo{year}{2003}).

\bibitem[{\citenamefont{{Lawley}}(2020{\natexlab{b}})}]{lawley2020}
\bibinfo{author}{\bibfnamefont{S.~D.} \bibnamefont{{Lawley}}},
  \bibinfo{journal}{Journal of Mathematical Biology}
  \textbf{\bibinfo{volume}{80}}, \bibinfo{pages}{2301}
  (\bibinfo{year}{2020}{\natexlab{b}}).

\bibitem[{\citenamefont{{Grebenkov} et~al.}(2020)\citenamefont{{Grebenkov},
  {Metzler}, and {Oshanin}}}]{2020NJPh...22j3004G}
\bibinfo{author}{\bibfnamefont{D.~S.} \bibnamefont{{Grebenkov}}},
  \bibinfo{author}{\bibfnamefont{R.}~\bibnamefont{{Metzler}}},
  \bibnamefont{and}
  \bibinfo{author}{\bibfnamefont{G.}~\bibnamefont{{Oshanin}}},
  \bibinfo{journal}{New Journal of Physics} \textbf{\bibinfo{volume}{22}},
  \bibinfo{eid}{103004} (\bibinfo{year}{2020}).

\bibitem[{\citenamefont{{Sawford}}(2001)}]{2001AnRFM..33..289S}
\bibinfo{author}{\bibfnamefont{B.}~\bibnamefont{{Sawford}}},
  \bibinfo{journal}{Annual Review of Fluid Mechanics}
  \textbf{\bibinfo{volume}{33}}, \bibinfo{pages}{289} (\bibinfo{year}{2001}).

\bibitem[{\citenamefont{{Bourgoin} et~al.}(2006)\citenamefont{{Bourgoin},
  {Ouellette}, {Xu}, {Berg}, and {Bodenschatz}}}]{2006Sci...311..835B}
\bibinfo{author}{\bibfnamefont{M.}~\bibnamefont{{Bourgoin}}},
  \bibinfo{author}{\bibfnamefont{N.~T.} \bibnamefont{{Ouellette}}},
  \bibinfo{author}{\bibfnamefont{H.}~\bibnamefont{{Xu}}},
  \bibinfo{author}{\bibfnamefont{J.}~\bibnamefont{{Berg}}}, \bibnamefont{and}
  \bibinfo{author}{\bibfnamefont{E.}~\bibnamefont{{Bodenschatz}}},
  \bibinfo{journal}{Science} \textbf{\bibinfo{volume}{311}},
  \bibinfo{pages}{835} (\bibinfo{year}{2006}).

\bibitem[{\citenamefont{{Salazar} and {Collins}}(2009)}]{2009AnRFM..41..405S}
\bibinfo{author}{\bibfnamefont{J.~P.~L.~C.} \bibnamefont{{Salazar}}}
  \bibnamefont{and} \bibinfo{author}{\bibfnamefont{L.~R.}
  \bibnamefont{{Collins}}}, \bibinfo{journal}{Annual Review of Fluid Mechanics}
  \textbf{\bibinfo{volume}{41}}, \bibinfo{pages}{405} (\bibinfo{year}{2009}).

\bibitem[{\citenamefont{{Rast} and {Pinton}}(2011)}]{2011PhRvL.107u4501R}
\bibinfo{author}{\bibfnamefont{M.~P.} \bibnamefont{{Rast}}} \bibnamefont{and}
  \bibinfo{author}{\bibfnamefont{J.-F.} \bibnamefont{{Pinton}}},
  \bibinfo{journal}{\prl} \textbf{\bibinfo{volume}{107}}, \bibinfo{eid}{214501}
  (\bibinfo{year}{2011}).

\bibitem[{\citenamefont{{Bourgoin}}(2015)}]{2015JFM...772..678B}
\bibinfo{author}{\bibfnamefont{M.}~\bibnamefont{{Bourgoin}}},
  \bibinfo{journal}{Journal of Fluid Mechanics} \textbf{\bibinfo{volume}{772}},
  \bibinfo{pages}{678} (\bibinfo{year}{2015}).

\bibitem[{\citenamefont{{Rast} et~al.}(2016)\citenamefont{{Rast}, {Pinton}, and
  {Mininni}}}]{2016PhRvE..93d3120R}
\bibinfo{author}{\bibfnamefont{M.~P.} \bibnamefont{{Rast}}},
  \bibinfo{author}{\bibfnamefont{J.-F.} \bibnamefont{{Pinton}}},
  \bibnamefont{and} \bibinfo{author}{\bibfnamefont{P.~D.}
  \bibnamefont{{Mininni}}}, \bibinfo{journal}{\pre}
  \textbf{\bibinfo{volume}{93}}, \bibinfo{eid}{043120} (\bibinfo{year}{2016}).

\bibitem[{\citenamefont{{Rice}}(1945)}]{1945BSTJ...24...46R}
\bibinfo{author}{\bibfnamefont{S.~O.} \bibnamefont{{Rice}}},
  \bibinfo{journal}{Bell System Technical Journal}
  \textbf{\bibinfo{volume}{24}}, \bibinfo{pages}{46} (\bibinfo{year}{1945}).

\bibitem[{\citenamefont{Cole et~al.}(2011)\citenamefont{Cole, Beck,
  Haji-Sheikh, and Litkouhi}}]{cole2010heat}
\bibinfo{author}{\bibfnamefont{K.~D.} \bibnamefont{Cole}},
  \bibinfo{author}{\bibfnamefont{J.~V.} \bibnamefont{Beck}},
  \bibinfo{author}{\bibfnamefont{A.}~\bibnamefont{Haji-Sheikh}},
  \bibnamefont{and} \bibinfo{author}{\bibfnamefont{B.}~\bibnamefont{Litkouhi}},
  \emph{\bibinfo{title}{Heat Conduction Using Green's Functions}}, Series in
  Computational Methods and Physical Processes in Mechanics and Thermal
  Sciences (\bibinfo{publisher}{CRC Press, Taylor \& Francis Group},
  \bibinfo{year}{2011}).

\bibitem[{\citenamefont{{Agrawal} et~al.}(2018)\citenamefont{{Agrawal}, {Rast},
  {Go{\v{s}}i{\'c}}, {Bellot Rubio}, and {Rempel}}}]{2018ApJ...854..118A}
\bibinfo{author}{\bibfnamefont{P.}~\bibnamefont{{Agrawal}}},
  \bibinfo{author}{\bibfnamefont{M.~P.} \bibnamefont{{Rast}}},
  \bibinfo{author}{\bibfnamefont{M.}~\bibnamefont{{Go{\v{s}}i{\'c}}}},
  \bibinfo{author}{\bibfnamefont{L.~R.} \bibnamefont{{Bellot Rubio}}},
  \bibnamefont{and} \bibinfo{author}{\bibfnamefont{M.}~\bibnamefont{{Rempel}}},
  \bibinfo{journal}{\apj} \textbf{\bibinfo{volume}{854}}, \bibinfo{eid}{118}
  (\bibinfo{year}{2018}).

\bibitem[{\citenamefont{{Abramowitz} and
  {Stegun}}(1972)}]{1972NBSAbramowitzandStegun}
\bibinfo{author}{\bibfnamefont{M.}~\bibnamefont{{Abramowitz}}}
  \bibnamefont{and} \bibinfo{author}{\bibfnamefont{I.~A.}
  \bibnamefont{{Stegun}}}, \emph{\bibinfo{title}{{Handbook of mathematical
  functions : With formulas, graphs, and mathematical tables}}}
  (\bibinfo{publisher}{US Department of Commerce, National Bureau of Standards,
  Applied Mathematical Series, 55}, \bibinfo{year}{1972}).

\bibitem[{\citenamefont{{Carslaw} and {Jaeger}}(1959)}]{1959chs..book.....C}
\bibinfo{author}{\bibfnamefont{H.~S.} \bibnamefont{{Carslaw}}}
  \bibnamefont{and} \bibinfo{author}{\bibfnamefont{J.~C.}
  \bibnamefont{{Jaeger}}}, \emph{\bibinfo{title}{{Conduction of heat in
  solids}}} (\bibinfo{publisher}{Oxford University Press},
  \bibinfo{year}{1959}).

\bibitem[{\citenamefont{{Cao} and {West}}(1997)}]{caoandwest1997}
\bibinfo{author}{\bibfnamefont{G.}~\bibnamefont{{Cao}}} \bibnamefont{and}
  \bibinfo{author}{\bibfnamefont{M.}~\bibnamefont{{West}}},
  \bibinfo{journal}{Communications in Statistics - Theory and Methods}
  \textbf{\bibinfo{volume}{26}}, \bibinfo{pages}{755} (\bibinfo{year}{1997}).

\bibitem[{\citenamefont{Philip}(2007)}]{jphilip2007}
\bibinfo{author}{\bibfnamefont{J.}~\bibnamefont{Philip}},
  \bibinfo{journal}{TRITA MAT} \textbf{\bibinfo{volume}{7}}
  (\bibinfo{year}{2007}).

\bibitem[{\citenamefont{Taboga}(2012)}]{taboga2012lectures}
\bibinfo{author}{\bibfnamefont{M.}~\bibnamefont{Taboga}},
  \emph{\bibinfo{title}{Lectures on Probability Theory and Mathematical
  Statistics}} (\bibinfo{publisher}{CreateSpace Independent Publishing
  Platform}, \bibinfo{year}{2012}).

\bibitem[{\citenamefont{Winkelmann}(1995)}]{winkelmann1995}
\bibinfo{author}{\bibfnamefont{R.}~\bibnamefont{Winkelmann}},
  \bibinfo{journal}{Journal of Business \& Economic Statistics}
  \textbf{\bibinfo{volume}{13}}, \bibinfo{pages}{467} (\bibinfo{year}{1995}).

\bibitem[{\citenamefont{{Bertini} et~al.}(2015)\citenamefont{{Bertini}, {De
  Sole}, {Gabrielli}, {Jona-Lasinio}, and {Landim}}}]{2015RvMP...87..593B}
\bibinfo{author}{\bibfnamefont{L.}~\bibnamefont{{Bertini}}},
  \bibinfo{author}{\bibfnamefont{A.}~\bibnamefont{{De Sole}}},
  \bibinfo{author}{\bibfnamefont{D.}~\bibnamefont{{Gabrielli}}},
  \bibinfo{author}{\bibfnamefont{G.}~\bibnamefont{{Jona-Lasinio}}},
  \bibnamefont{and} \bibinfo{author}{\bibfnamefont{C.}~\bibnamefont{{Landim}}},
  \bibinfo{journal}{Reviews of Modern Physics} \textbf{\bibinfo{volume}{87}},
  \bibinfo{pages}{593} (\bibinfo{year}{2015}), \eprint{1404.6466}.

\bibitem[{\citenamefont{{Agranov} et~al.}(2019)\citenamefont{{Agranov},
  {Krapivsky}, and {Meerson}}}]{2019PhRvE..99e2102A}
\bibinfo{author}{\bibfnamefont{T.}~\bibnamefont{{Agranov}}},
  \bibinfo{author}{\bibfnamefont{P.~L.} \bibnamefont{{Krapivsky}}},
  \bibnamefont{and}
  \bibinfo{author}{\bibfnamefont{B.}~\bibnamefont{{Meerson}}},
  \bibinfo{journal}{\pre} \textbf{\bibinfo{volume}{99}}, \bibinfo{eid}{052102}
  (\bibinfo{year}{2019}).

\bibitem[{\citenamefont{{Agranov} and {Meerson}}(2018)}]{2018PhRvL.120l0601A}
\bibinfo{author}{\bibfnamefont{T.}~\bibnamefont{{Agranov}}} \bibnamefont{and}
  \bibinfo{author}{\bibfnamefont{B.}~\bibnamefont{{Meerson}}},
  \bibinfo{journal}{\prl} \textbf{\bibinfo{volume}{120}}, \bibinfo{eid}{120601}
  (\bibinfo{year}{2018}).

\bibitem[{\citenamefont{{Grebenkov}}(2019{\natexlab{b}})}]{2019PhRvE.100f2110G}
\bibinfo{author}{\bibfnamefont{D.~S.} \bibnamefont{{Grebenkov}}},
  \bibinfo{journal}{\pre} \textbf{\bibinfo{volume}{100}}, \bibinfo{eid}{062110}
  (\bibinfo{year}{2019}{\natexlab{b}}).

\bibitem[{\citenamefont{{Grebenkov}}(2020{\natexlab{b}})}]{2020JSMTE2020j3205G}
\bibinfo{author}{\bibfnamefont{D.~S.} \bibnamefont{{Grebenkov}}},
  \bibinfo{journal}{Journal of Statistical Mechanics: Theory and Experiment}
  \textbf{\bibinfo{volume}{2020}}, \bibinfo{eid}{103205}
  (\bibinfo{year}{2020}{\natexlab{b}}).

\bibitem[{\citenamefont{{Grebenkov}}(2021)}]{2021JPhA...54a5003G}
\bibinfo{author}{\bibfnamefont{D.~S.} \bibnamefont{{Grebenkov}}},
  \bibinfo{journal}{Journal of Physics A Mathematical General}
  \textbf{\bibinfo{volume}{54}}, \bibinfo{eid}{015003} (\bibinfo{year}{2021}).

\bibitem[{\citenamefont{Warren}(2018)}]{doi:10.1177/0963721417746743}
\bibinfo{author}{\bibfnamefont{W.~H.} \bibnamefont{Warren}},
  \bibinfo{journal}{Current Directions in Psychological Science}
  \textbf{\bibinfo{volume}{27}}, \bibinfo{pages}{232} (\bibinfo{year}{2018}).

\end{thebibliography}

\end{document}